\documentclass[]{aa} %
\usepackage{graphicx}
\usepackage{txfonts}
\usepackage{verbatim}
\usepackage{bm}

\begin{document}

%
   \title{The NIKA2 large-field-of-view millimetre continuum camera for the 30m IRAM telescope}


\author
{
R.~Adam \inst{\ref{OCA}}
\and A.~Adane \inst{\ref{USTHB}}
\and  P.A.R.~Ade \inst{\ref{Cardiff}}
\and  P.~Andr\'e \inst{\ref{CEA}}
\and  A.~Andrianasolo \inst{\ref{IPAG}} %
\and  H.~Aussel \inst{\ref{CEA}}
\and  A.~Beelen \inst{\ref{LAM}}
\and  A.~Beno\^it \inst{\ref{Neel}}
\and  A.~Bideaud \inst{\ref{Neel}}
\and  N.~Billot \inst{\ref{IRAME}}
\and  O.~Bourrion \inst{\ref{LPSC}}
\and  A.~Bracco \inst{\ref{CEA}} %
\and  M.~Calvo \inst{\ref{Neel}}
\and  A.~Catalano \inst{\ref{LPSC}}
\and  G.~Coiffard \inst{\ref{IRAMF}} %
\and  B.~Comis \inst{\ref{LPSC}}
\and  M. De Petris \inst{\ref{Roma}} %
\and  F.-X.~D\'esert \inst{\ref{IPAG}}
\and  S.~Doyle \inst{\ref{Cardiff}}
\and  E.F.C.~Driessen \inst{\ref{IRAMF}}
\and  R.~Evans \inst{\ref{Cardiff}} %
\and  J.~Goupy \inst{\ref{Neel}}
\and  C.~Kramer \inst{\ref{IRAME}}
\and  G.~Lagache \inst{\ref{LAM}}
\and  S.~Leclercq \inst{\ref{IRAMF}}
\and  J.-P.~Leggeri \inst{\ref{Neel}} %
\and  J.-F.~Lestrade \inst{\ref{LERMA}}
\and  J.F.~Mac\'ias-P\'erez \inst{\ref{LPSC}}
\and  P.~Mauskopf \inst{\ref{Arizona}}
\and  F.~Mayet \inst{\ref{LPSC}}
\and  A.~Maury \inst{\ref{CEA}}  %
\and  A.~Monfardini \inst{\ref{Neel}}\thanks{Corresponding author, e-mail: alessandro.monfardini@neel.cnrs.fr} 
\and S.~Navarro \inst{\ref{IRAME}} %
\and  E.~Pascale \inst{\ref{Cardiff}}
\and  L.~Perotto \inst{\ref{LPSC}}
\and  G.~Pisano \inst{\ref{Cardiff}}
\and  N.~Ponthieu \inst{\ref{IPAG}}
\and  V.~Rev\'eret \inst{\ref{CEA}}
\and  A.~Rigby \inst{\ref{Cardiff}} %
\and A.~Ritacco \inst{\ref{IRAME}}
\and  C.~Romero \inst{\ref{IRAMF}}
\and  H.~Roussel \inst{\ref{IAP}}
\and  F.~Ruppin \inst{\ref{LPSC}}
\and  K.~Schuster \inst{\ref{IRAMF}}
\and  A.~Sievers \inst{\ref{IRAME}}
\and  S.~Triqueneaux \inst{\ref{Neel}}
\and  C.~Tucker \inst{\ref{Cardiff}}
\and  R.~Zylka \inst{\ref{IRAMF}}
}

         \institute
        {
Laboratoire Lagrange, Universit\'e et Observatoire de la C\^ote d'Azur, CNRS, Blvd de l'Observatoire, F-06304 Nice, France\label{OCA}
\and
University of Sciences and Technology Houari Boumediene (U.S.T.H.B.), BP 32 El Alia, Bab Ezzouar 16111, Algiers, Algeria \label{USTHB}
\and
Astronomy Instrumentation Group, University of Cardiff, The Parade, CF24 3AA, United Kindgom\label{Cardiff}
\and
Laboratoire AIM, CEA/IRFU, CNRS/INSU, Universit\'e Paris Diderot, CEA-Saclay, 91191 Gif-Sur-Yvette, France\label{CEA}
\and
Univ. Grenoble Alpes, CNRS, IPAG, 38000 Grenoble, France\label{IPAG}
\and
Aix Marseille Universit\'e, CNRS, LAM (Laboratoire d'Astrophysique de Marseille), F-13388 Marseille, France \label{LAM}
\and
Institut N\'eel, CNRS and Universit\'e Grenoble Alpes (UGA), 25 av. des Martyrs, F-38042 Grenoble, France \label{Neel}
\and
Institut de RadioAstronomie Millim\'etrique (IRAM), Granada, Spain \label{IRAME}
\and
Laboratoire de Physique Subatomique et de Cosmologie, Universit\'e Grenoble Alpes, CNRS, 53, av. des Martyrs, Grenoble, France\label{LPSC}
\and
Institut de RadioAstronomie Millim\'etrique (IRAM), Grenoble, France \label{IRAMF}
\and
Dipartimento di Fisica, Universit\`a di Roma La Sapienza, Piazzale Aldo Moro 5, I-00185 Roma, Italy \label{Roma}
\and 
LERMA, CNRS, Observatoire de Paris, 61 avenue de l'Observatoire, Paris, France \label{LERMA}
\and
School of Earth and Space Exploration and Department of Physics, Arizona State University, Tempe, AZ 85287 \label{Arizona}
\and 
Institut d'Astrophysique de Paris, CNRS (UMR7095), 98 bis boulevard Arago, F-75014 Paris, France \label{IAP}
             }

    \authorrunning{Adam et. al.}

   \date{Received December XX, XXXX; accepted XXX XX, XXXX}


  \abstract
   {Millimetre-wave continuum astronomy is today an indispensable tool for both general astrophysics studies (e.g. star formation, nearby galaxies) and cosmology (e.g. CMB - cosmic microwave background and high-redshift galaxies). General purpose, large-field-of-view instruments are needed to map the sky at intermediate angular scales not accessible by the high-resolution interferometers (e.g. ALMA in Chile, NOEMA in the French Alps) and by the coarse angular resolution space-borne or ground-based surveys (e.g. Planck, ACT, SPT). These instruments have to be installed at the focal plane of the largest single-dish telescopes, which are placed at high altitude on selected dry observing sites. In this context, we have constructed and deployed a three-thousand-pixel dual-band (150\,GHz and\,260 GHz, respectively 2\,mm and 1.15\,mm wavelengths) camera to image an instantaneous circular field-of-view of 6.5\,arcminutes in diameter, and configurable to map the linear polarisation at 260\,GHz.}
   {First, we are providing a detailed description of this instrument, named NIKA2 (New IRAM KID Arrays 2), in particular focussing on the cryogenics, optics, focal plane arrays based on Kinetic Inductance Detectors (KID), and the readout electronics. The focal planes and part of the optics are cooled down to the nominal 150\,mK operating temperature by means of an ad-hoc dilution refrigerator. 
Secondly, we are presenting the performance measured on the sky during the commissioning runs that took place between October 2015 and April 2017 at the 30-meter IRAM (Institut of Millimetric Radio Astronomy) telescope at Pico Veleta, near Granada (Spain).}
   {We have targeted a number of astronomical sources. Starting from beam-maps on primary and secondary calibrators we have then gone to extended sources and faint objects. Both internal (electronic) and on-the-sky calibrations are applied. The general methods are described in the present paper.}
   {NIKA2 has been successfully deployed and commissioned, performing in-line with expectations. In particular, NIKA2 exhibits full width at half maximum (FWHM) angular resolutions of around 11 and 17.5 arc-seconds at respectively 260 and 150\,GHz. The  noise equivalent flux densities (NEFD) are, at these two respective frequencies, 33$\pm$2 and 8$\pm$1 mJy $\cdot\textrm{s}^{1/2}$. A first successful science verification run was achieved in April 2017. The instrument is currently offered to the astronomy community and will remain available for at least the following ten years.}
  {}


   \maketitle
%

\section{Introduction}

In the past decades progress of astronomical instruments, in particular the development of large arrays of background-limited detectors, has led to a golden era of millimetre and sub-millimetre continuum astronomy. A number of instruments operate hundreds to thousands of very sensitive pixels. The majority of these instruments, however, are designed to execute specific scientific programs, most likely related to the search of the primordial polarisation modes in the Cosmic Microwave Background (CMB). Few general purpose platforms, like the one described in this paper, are currently available to the general astronomy community. Among them, we cite, for example, Artemis (\cite{Reveret2014}) and LABOCA (\cite{Siringo2009}) on APEX (Chile), SCUBA2 (\cite{Holland2013}) on JCMT (Hawaii), AzTEC (\cite{Chavez-Dagostino2016}) on the LMT (Mexico), MUSTANG2 (\cite{Dicker2014}) on GBT (USA), and HAWC+ (\cite{Staguhn2016}) on-board SOFIA. These cameras are all based on the classical bolometric detection principle, so far the best approach for this application. In the past ten years, the Kinetic Inductance Detectors (KID) concurrent technology has demonstrated its competitiveness. For example, the pathfinder NIKA instrument at the IRAM 30-meter telescope, equipped with 356 pixels split over two arrays, has demonstrated state-of-the art performance in terms of sensitivity, stability, and dynamic range (\cite{Monfardini2010, Monfardini2011, Catalano2014, Adam2014}). The most recent advancements in the instrumental domain are described in detail in the LTD16 (Low Temperature Detectors 16) workshop proceedings \footnote{LTD16, Grenoble, July $20-24^{th}$ 2015,  Journal of Low Temperature Physics 184, numbers 1/2 and 3/4, 2016.}.

Despite the spectacular progression of the technology, sub-millimetre and millimetre studies are often limited by the mapping speed of intermediate resolution, that is, large single dishes, instruments and their spectral coverage. 

Concerning Galactic studies, deep millimetre and sub-millimetre observations at high angular resolution, in intensity and in polarisation, are needed to better understand how star formation proceeds in the interstellar medium (ISM). Solar-type stars form within regions of cold gas in the ISM. These molecular clouds are characterised by an intricate filamentary structure of matter, which hosts the progenitors of stars, that is, pre-stellar and proto-stellar cores (\cite{Andre2010}, \cite{Konyves2015}, \cite{Bracco2017}).

A second key subject is represented by the study of nearby galaxies which aims at separating physical components by dissentangling the emission (e.g. thermal dust, free-free, or synchrotron). This allows one to precisely measure the star formation rate in different environments and regions. In this framework, the large instantaneous field-of-view of NIKA2 is an asset.

On the cosmological side, existing CMB experiments have proven to be very efficient in detecting clusters of galaxies via the Sunyaev-Zel\textquoteright dovich (SZ) effect (\cite{plancksz2,actsz,sptsz}) and have provided the best cluster cosmological results to date (\cite{plancksp2,plancknc2}). However, their poor angular resolution limits the cosmological interpretation of the data and in particular the study of the complex intra cluster medium (ICM) physics, which may bias the observable to cluster-mass scaling relation (\cite{plancknc2}). This bias might be of particular importance for high-redshift, that is, early-stage, galaxy clusters. The dual-band capability of NIKA2 is, in this case, crucial.

Similarly, distant universe studies via deep surveys will benefit from a large instantaneous field-of-view and sensitivity to cover sky regions at the confusion limit. This results in detecting dust-obscured optically faint galaxies during their major episodes of star formation in the early universe (\cite{Bethermin2017}, \cite{Geach2017}). 

NIKA2 and the IRAM 30-meter telescope represent today ideal tools to address these scientific questions, and many others. The fundamental characteristics of NIKA2 are dual color and polarisation capabilities, high sensitivity, high angular resolution and an instantaneous field-of-view of 6.5 arc-minutes. Besides the intrinsic scientific impact, NIKA2 represents the first demonstration of competitive performance using large-format (i.e. thousands of pixels) Kinetic Inductance Detector (KID (\cite{Day2003}, \cite{Doyle2010}) cameras operating at millimetre or sub-millimetre wavelengths.


In Section \ref{The NIKA2 Instrument} we describe the overall instrument design, including cryogenics, focal planes, optics and readout electronics. In Section \ref{Measurement principle} we discuss the observing methods, concentrating in particular on the photometric calibration procedures. We then, in Sections \ref{Observations and performance} and \ref{Illustration of NIKA2 mapping capabilities}, present the results from the intensity commissioning runs at the 30-meter telescope. 


\section{The NIKA2 Instrument}
\label{The NIKA2 Instrument}

NIKA2 is a multi-purpose tool able to simultaneously image a field-of-view of 6.5\,arcminutes in diameter at 150 and 260\,GHz. When run in polarimetric mode, it maps the linear polarisation at 260\,GHz. In order not to degrade the native angular resolution of the 30-meter telescope, and at the same time cover a large field-of-view, it employs a total of around 2,900\,detectors split over three distinct monolithic arrays of KID. 
In this Section, we describe the main instrument sub-systems and the in-laboratory characterisation procedures.

 \subsection{The cryostat}

\begin{figure}[h]
   \centering
   \includegraphics[width=.95\linewidth]{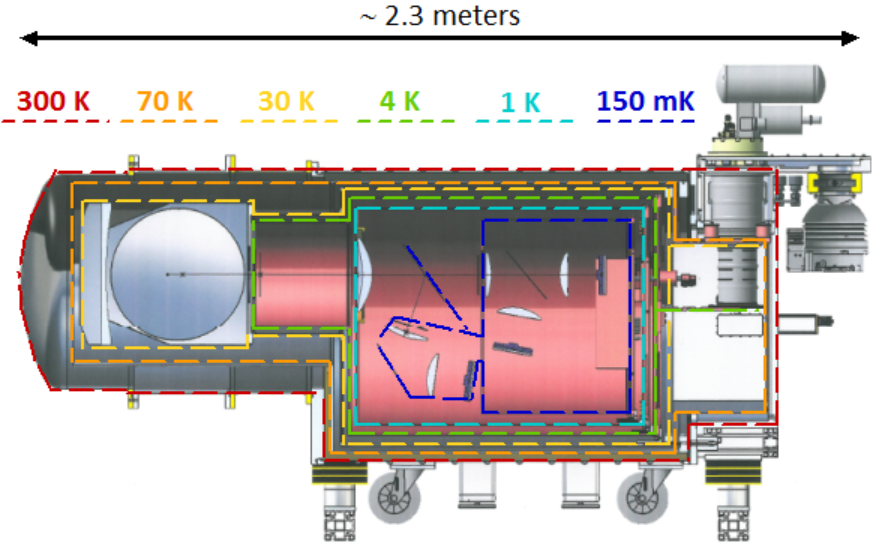}
      \caption{(Colour online) Cross-section of the NIKA2 instrument illustrating the different cryogenic stages. The total weight of the cryostat is close to 1,300\,kg. The 150\,mK section includes the arrays, the dichroic, the polariser and five high-density polyethilene (HDPE) lenses.}
         \label{Cryostat_cryo}
\end{figure}

In order to ensure optimal operation of the detectors and minimise the in-band parasitic radiation, the focal plane arrays, and the last portion of the optics, are cooled down to a base temperature of around 150\,mK by means of a dilution fridge. The base temperature must, in fact, be roughly one order of magnitude lower than the thin-aluminium superconducting critical temperature. The home-made dilution insert is completely independent and compatible with any cryostat providing a stable 4\,K temperature input and suitable mechanical and fluid attachment points. We stress that no recycling is needed, the hold time being, in principle, infinite. The dilution refrigerator, and the rest of the cryostat, has been entirely designed and realised by CNRS Grenoble. NIKA2 employs two Cryomech PT415 pulses-tubes, each delivering a cooling power of 1.35\,W at the reference temperature of 4.2\,K (second stage) and several tens of Watts at 30-70\,K (first stage). The base temperature of these machines is of the order of 3\,K, sufficient to start the isotopic dilution process. The large cooling power available on the pulse tube's first stages allows the integration of a part of the optics baffles at temperatures between 4 and 30\,K within the cryostat. A cross-section of the cryostat is shown in Fig.~\ref{Cryostat_cryo} to illustrate the different cryogenic stages.

Gas heat exchangers are adopted at both pulse-tubes stages to ensure good thermal contact, avoiding at the same time direct mechanical contact between the vibrating parts and the sensitive inner components, that is, the detectors and cold electronics. An external mechanical regulation of the pulse-tube positions allows the optimisation of the cooling power and at the same time the minimisation of the shaking of the coldest components. 

The whole instrument is made of thousands of mechanical pieces, with a total weight of around 1.3\,tons when assembled. The weight of the 150\,mK stage is of the order of 100\,kg, including several kg of high-density polyethilene (HDPE) low-conductance lenses. Radiation screens are placed at 1\,K (still of the dilution fridge), 4-8\,K (pulse-tube second stages), 30\,K and 70\,K (pulse-tube first stages).

Selected inner parts, at each stage of temperature, are coated with a high-emissivity mixture of black STYCAST 2850, SiC grains, and carbon powder. This coating has demonstrated its effectiveness at millimetre wavelengths in suppressing unwanted reflections (\cite{Calvo2010}).

Following the experience accumulated with NIKA, magnetic shielding is added on each cryogenic stage, employing high-permittivity materials down to 1\,K, and a pure aluminium superconducting screen cooled to 150\,mK enclosing the detectors. The screening is needed in order to suppress: a) the Earth's magnetic field and its variations, in the instrument reference frame, during the telescope slews in azimuth; and b) the magnetic field variations induced by the antenna moving in elevation.  

The operation of NIKA2 does not require external cryogenic liquids. Even the needed cold traps are included in the cryostat and cooled by the pulse-tubes themselves. The cleaned helium isotopic mixture is condensed into liquid when the system is cold, but it remains in a closed circuit. The cryostat and related sub-systems are fully remotely controlled. The whole cool-down procedure, largely automated, lasts about five days. Four days are required for the pre-cooling and thermalisation of the three coldest stages (see Fig.~\ref{Cryostat_cryo}) at around 4\,K. During the last 24 hours, the dilution procedure is started, allowing the further cooling down to base temperature. Two additional days before, stable observations are usually foreseen in order to ensure the perfect thermalisation of all the low-thermal-conductance optics elements, such as the lenses, filters and baffle coating. The system is designed for continuous operations and long observational runs. So far the base temperature has shown the required stability over roughly one month, with no signs of degradation in performance. The stability of the detector's temperature is better than 0.1\,mK RMS over the duration of a typical observational block (scan), that is, roughly 15\,min. The common-mode effect on the detectors of such temperature variations is in this way at least one order of magnitude lower than the atmospheric fluctuations, even assuming the very best observing conditions. We stress in this context that the maximum gradient between the coldest cryostat parts and the farthest KID array is around 20\,mK.

 \subsection{The focal plane arrays}

Each array is fabricated on a single 4" high-resistivity silicon wafer, on which an aluminium film (t = 18\,nm) is deposited by e-beam evaporation under ultra-high vacuum conditions. The use of thin superconducting films has a double advantage. First, it increases the kinetic inductance of the strip, making the detectors more responsive, and second, it allows, through its normal state sheet resistance, an almost perfect match of the Lumped Element Kinetic Inductance Detector ( LEKID) meander to the free space impedance of the incoming wave, ensuring a quantum efficiency exceeding 90\% at the peak. The NIKA2 pixels are all based on the Hilbert fractal geometry that we proposed some years ago (\cite{Roesch2012}). 

In NIKA, we adopted more classical pixels coupled to a co-planar waveguide (CPW) readout line, with wire bonds across the central line to suppress the spurious slotline mode. Purely microfabricated bridges were developed as well. The slotline mode is associated to a symmetry-breaking between the ground planes on both sides with the central strip. To optimise the optical coupling to the incoming millimetre radiation, we adopted a back-illumination configuration, in which the light passes through the silicon wafer before reaching the pixels. To attenuate the refraction index mismatch, we micro-machined a grid of perpendicular grooves on the back-side of the wafer, resulting in an effective dielectric constant which is in between vacuum and silicon (\cite{Goupy2016}). The total thickness of the silicon wafer, and the depth of the grooves, were chosen to optimise the anti-reflection effect. To maximise the in-band radiation absorption, a superconducting lid was then set at an optimised distance behind the detector plane, as a $\lambda/4$ backshort. 

The same approach was originally planned for NIKA2. During the phase of the detector's development, however, we realised the practical limitations of the CPW coupling approach, in particular considering the thousands of bonds required to ensure the exclusive propagation of the CPW mode. We then decided to study and optimise a different kind of transmission line, the microstrip (MS). This kind of feedline only supports one propagating mode, and is thus immune to the risk of spurious modes. Furthermore, the aluminium ground plane is located on the opposite side of the wafer with respect to the detectors. This might reduce the still poorly understood residual electro-magnetic cross-coupling between resonators (pixels).

\begin{figure}[h]
   \centering
    \includegraphics[width=.95\linewidth]{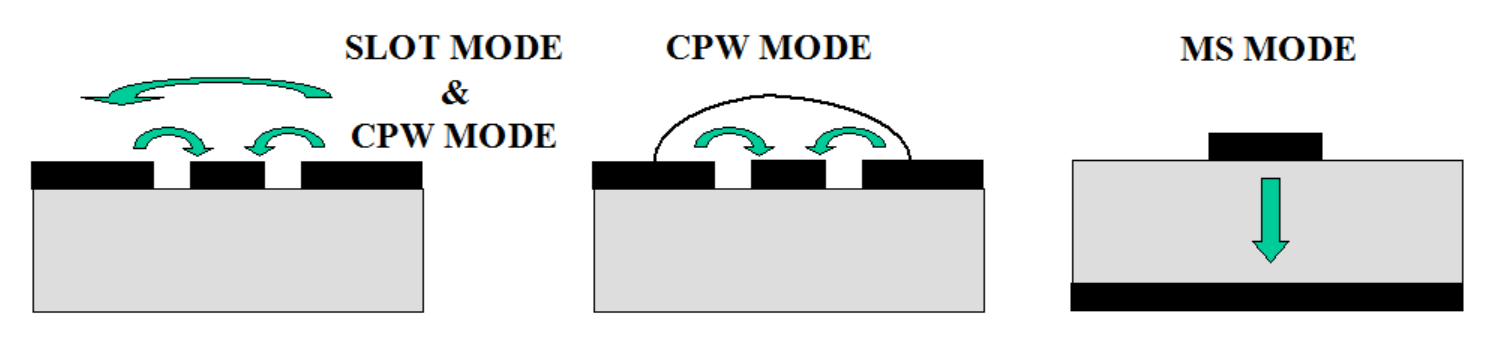}
      \caption{Schematic cut through a KID substrate. In grey, the high-resistance silicon wafer, while in black, the aluminium films are represented. The green arrows illustrate the direction of the electric field. \underline{Left:} the co-planar waveguide (CPW) transmission line without across-the-line bondings, associated to strongly non-uniform performance of the detector array. \underline{Centre:} CPW with across-the-line bondings, configuration adopted in NIKA. \underline{Right:} the microstrip (MS) configuration adopted in NIKA2, ensuring single-mode propagation and easiest implementation of very large arrays.}
         \label{CPWvsMS}
\end{figure}

The MS propagation mode shows an electric field oscillating in the dielectric substrate, between the strip line (feedline) and the underlying ground plane. This is illustrated in Fig.~\ref{CPWvsMS}. The main drawback of the MS coupling lies in the fact that it forces, at least for dual-polarisation imaging applications, front illumination of the detectors. It is thus more adapted for relatively narrow-band (e.g. $\Delta f / f  \leq 30 \%$) applications. This is however perfectly compatible with the NIKA2 goals.  

\begin{figure}[h]
   \centering
        \includegraphics[width=.95\linewidth]{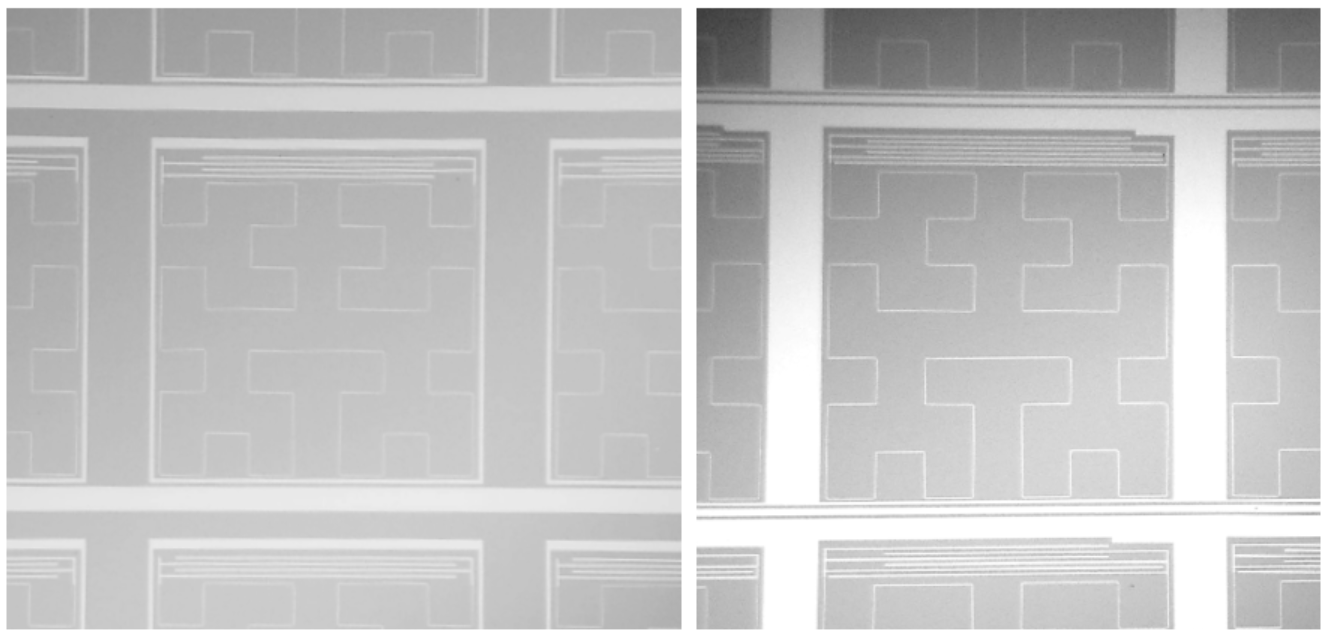}
     \caption{\underline{Left:} a front-illuminated microstrip (MS) pixel for the 260\,GHz band of NIKA2. The pixel size is $2\times2\textrm{\,mm}^2$. \underline{Right:} a back-illuminated coplanar waveguide (CPW) pixel used for the 150 GHz band in NIKA. The pixel size was in that case $2.3\times2.3\textrm{\,mm}^2$. Both designs are based on Hilbert-shape absorbers/inductors.}
         \label{Pixels}
\end{figure}

In both cases (CPW and MS), the distance between the pixels and the feedline is chosen in order to satisfy optimal coupling conditions. These are achieved when the coupling quality factor, $Q_c$, is of the same order as the internal quality factor $Q_i$ observed under typical loading conditions. In the case of NIKA2, we found an optimum at $Q_c\sim10,000$. A metal loop is added around each MS-coupled pixel to shield them from the feedline and achieve the wanted coupling without compromising the compactness of the pixel packaging (see Fig.~\ref{Pixels}). 

In NIKA2, the 150\,GHz channel is equipped with an array of 616\,pixels, arranged to cover a 78\,mm diameter circle. Each pixel has a size of $2.8\times2.8\textrm{\,mm}^2$. This is the maximum pixel size that can be adopted without significantly degrading the theoretical telescope resolution, as it corresponds roughly to a $1 F \lambda$ sampling of the focal plane. The array is connected over four different readout lines, and shows resonance frequencies between 0.9 and 1.4\,GHz. The thickness of the silicon substrate is around 150\,microns, to ensure a maximal optical absorption at 150\,GHz. 

\begin{figure}[h]
   \centering
    \includegraphics[width=.95\linewidth]{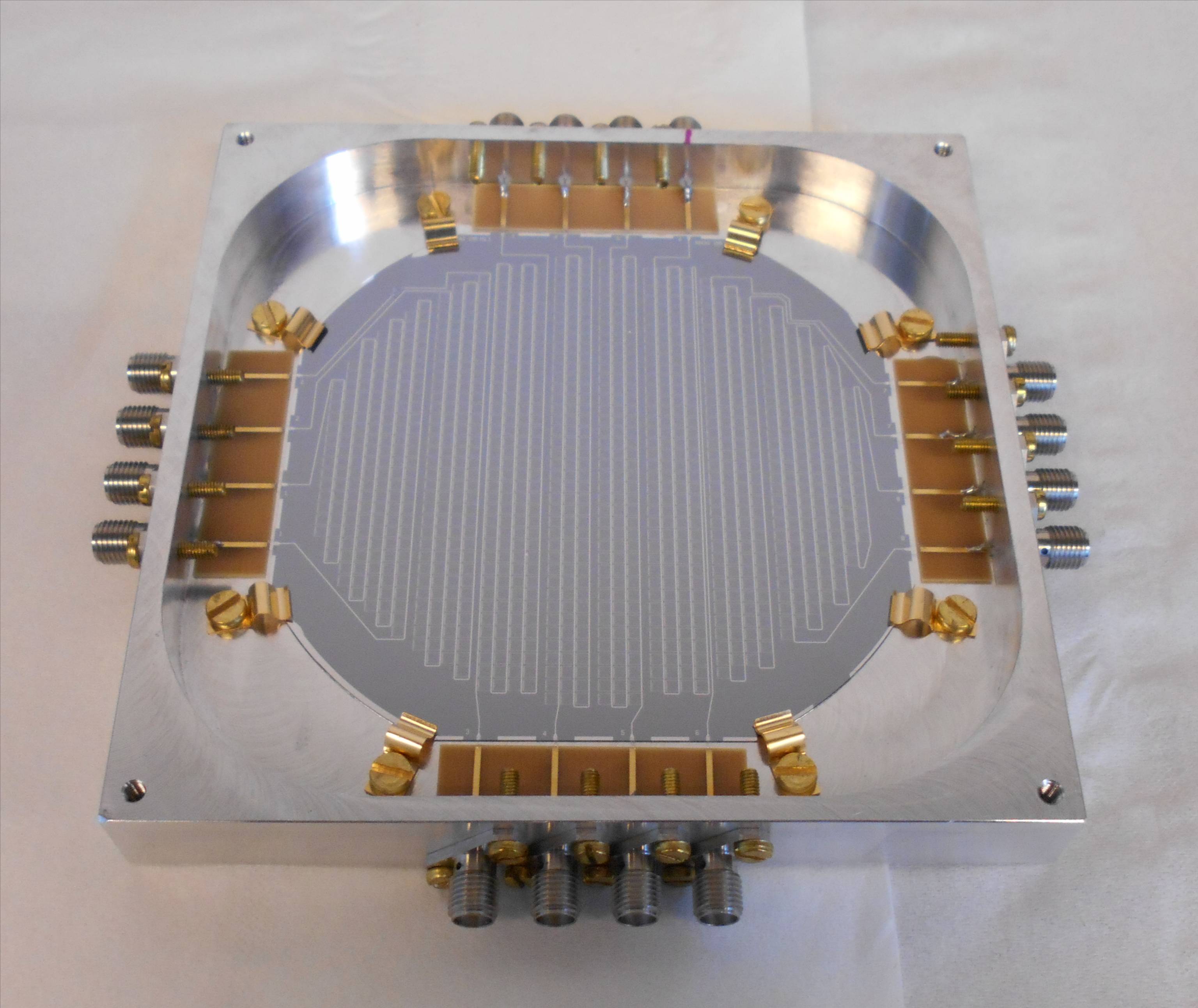}
      \caption{(Colour online) One of the 260\,GHz NIKA2 arrays after packaging. The number of pixels designed for this array is 1,140, connected via eight feed-lines and 16 SMA (SubMiniature version A) connectors to the external circuit. The front of the wafer can be seen here.}
         \label{Array}
\end{figure}

In the case of the 260\,GHz band detectors, the pixel size is $2\times 2\mathrm{\,mm}^2$, to ensure a comparable $1 F \lambda$ sampling of the focal plane. In order to fill the two 260\,GHz arrays, a total of 1,140 pixels are needed in each of them. The smaller pixel dimensions compared to the 150\,GHz band lead to slightly higher resonance frequencies that lie between 1.9 and 2.4\,GHz. Each of the 260\,GHz arrays is connected over eight different readout lines. The thickness of the substrate is 260\,microns, which maximises the optical absorption at 260\,GHz. A picture of one of the actual 260\,GHz arrays mounted in NIKA2 is shown in Fig.~\ref{Array}.

\begin{figure}[h]
   \centering
   \includegraphics[width=.95\linewidth]{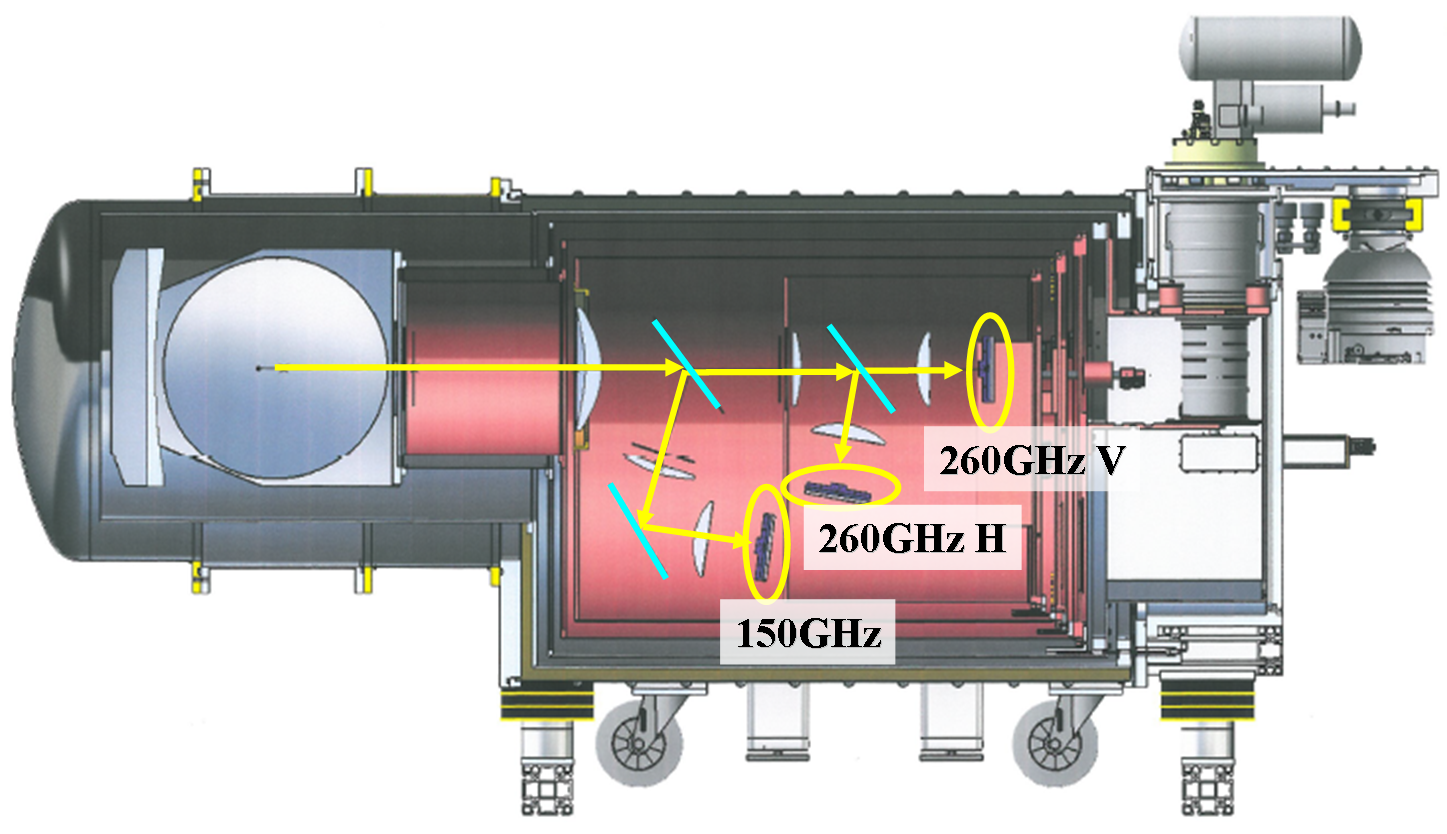}
      \caption{(Colour online) Cross-section of the NIKA2 instrument illustrating the position of the three detector arrays (150\,GHz, 260\,GHz-V and 260\,GHz-H). The optical axis and the photon direction of propagation are shown as well.}
         \label{Cryostat}
\end{figure}

We show in Fig.~\ref{Cryostat} an illustration of the positioning of the three arrays in the NIKA2 cryostat.
 
 \subsection{The cold optics}

In this Section we describe the internal (cooled) optics. More details concerning the telescope interface (room temperature) mirrors are given in Section~\ref{The integration at the telescope}.

NIKA2 is equipped with a reflective cold optics stage held at a temperature of around 30\,K. The two shaped mirrors (M7 and M8) are mounted in a specifically designed low-reflectance optical box in the cryostat nose. The stray-light suppression is further enhanced by a multi-stage baffle at 4\,K. The cold aperture stops, at a temperature of 150\,mK, are conservatively designed to be conjugated to the inner 27.5 metres of the primary mirror M1.

\begin{figure}[h]
   \centering
   \includegraphics[width=.95\linewidth]{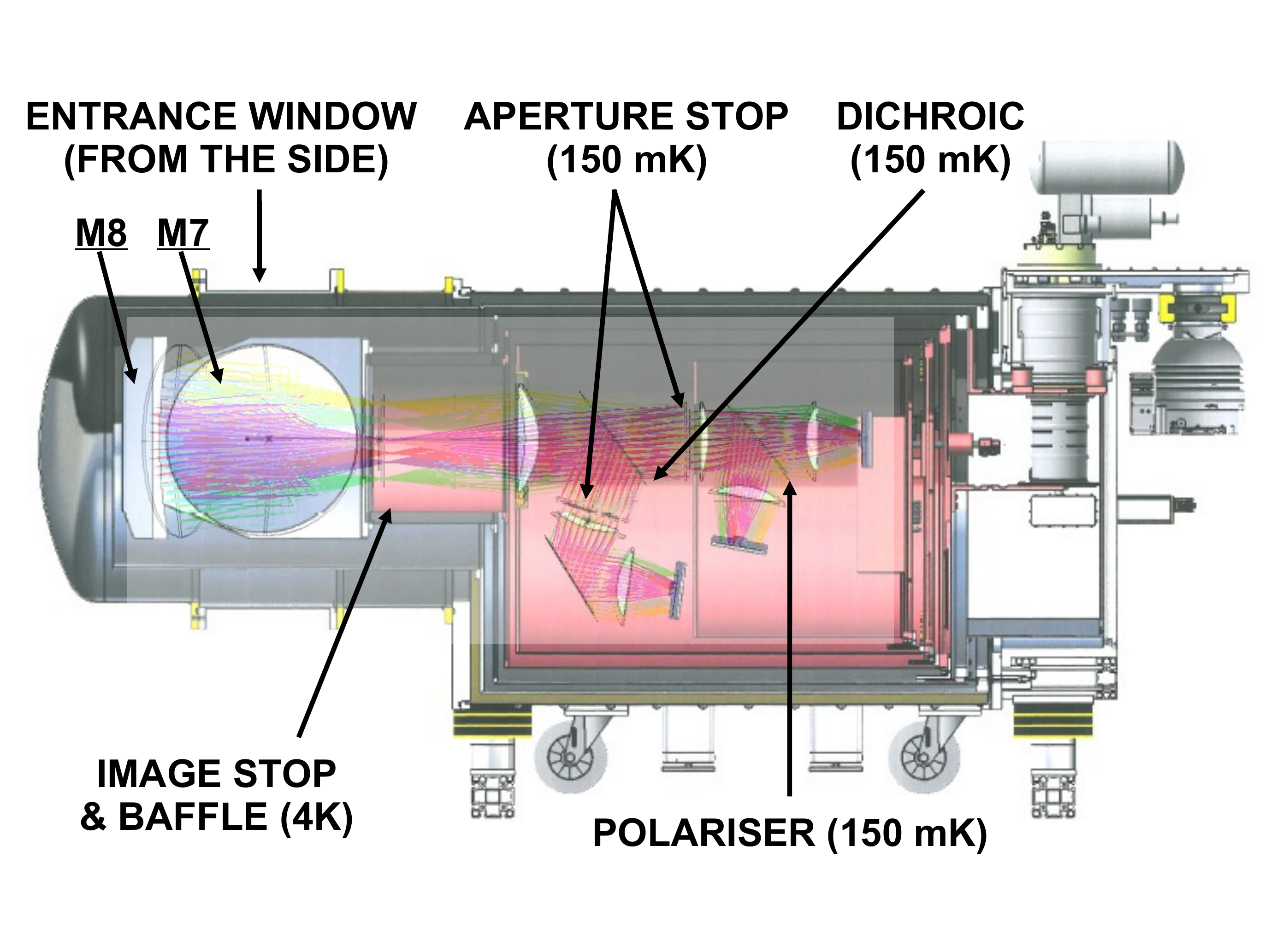}
      \caption{(Colour online) Cross-section of the NIKA2 instrument illustrating the cold optics and the main elements and surfaces described in the text. The cold mirrors M7 and M8 are mounted in the cryostat "nose" on the left side of the Figure.}
         \label{Cryostat_optics}
\end{figure}

The refractive elements of the NIKA2 cold optics are mounted at 1\,K and at the base temperature. The HDPE lenses, except for those placed in front of the 260\,GHz arrays, are anti-reflecting-coated. The coating is realised by a custom machining of the surfaces. A 30-centimeter-diameter air-gap dichroic splits the 150\,GHz (reflection) from the 260\,GHz (transmission) beams. This dichroic, ensuring that it is flatter relative to the standard hot-pressed ones, was developed in Cardiff specifically for NIKA2. A grid polariser ensures then the separation of the two linear polarisations on the 260\,GHz channel (V and H, see Fig.~\ref{Cryostat}). We refer to Fig.~\ref{Cryostat_optics} for a schematic cross-section of the inner optics.

The filtering of unwanted (off-band) radiation is provided by a suitable stack of multi-mesh filters placed at all temperature stages. In particular, three infrared-blocking filters are installed at 300\,K, 70\,K, and 30\,K. Multi-mesh low-pass filters, with decreasing cutoff frequencies, are mounted at 30\,K, 4\,K, 1\,K and the base temperature. Band-defining filters, custom-designed to optimally match the atmospheric windows (see Fig.~\ref{Fig4}), are installed at the base temperature. 

To exploit the NIKA2 polarisation capabilities, a modulator is added when operating the instrument in polarimetric mode; this consists of a multi-mesh hot-pressed half-wave-plate (HWP) (\cite{Pisano2008}) mounted, at room temperature, in front of the cryostat window. The modulator uses a stepping motor and is operated at mechanical frequencies of up to 3\,Hz, corresponding to a maximum of 12\,Hz on the effective polarisation modulation frequency. A similar setup was successfully used on NIKA (\cite{Ritacco2017}). In order to detect all the photons, the modulated polarised signal is then split onto the two 260\,GHz arrays by the 45 degree wire-grid polariser described above.

 \subsection{The readout electronics}
 \label{The readout electronics}

One of the key advantages of the KID technology is the simplicity of the cold electronics installed in the cryostat.
In NIKA2, each block of around 150 detectors is connected to single coaxial line providing the excitation at one end, and the readout at the other. The excitation lines, composed of stainless-steel cables, are running from 300\,K down to sub-Kelvin temperature. They are properly thermalised at each cryostat stage, and a fixed attenuation of 20\,dB is applied at 4\,K in order to suppress the room temperature thermal noise. Each excitation line ends with a SMA (SubMiniature version A) connector (EXCitation input) and an ad-hoc launcher connected, through superconducting (aluminium) micro-bonds, to the silicon wafer holding the detectors. The approximate excitation power per resonator is typically of the order of 10\,pW.

On the readout side, the same types of micro-bonds are used to transfer the signal out of the focal plane and to make it available on a second SMA connector (MEASurement output). Then a superconducting (Nb) coaxial cable is used to connect the measurement output directly to the input of a cryogenic low-noise amplifier (LNA). The amplified signal provided by the LNA is transferred through the remaining cryostat stages (up to 300\,K) via stainless-steel coaxial cables. The LNAs, which operate at frequencies up to 3\,GHz, show noise temperatures between 2\,K and 5\,K and are held at a physical temperature of about 8\,K. This means that the input amplifier noise is equivalent to the thermal Johnson noise of a 50-Ohm load placed between 2\,K and 5\,K . The cryogenic amplifiers used in NIKA2 were developed, fabricated, and tested at the Yebes observatory and TTI Norte company, both located in Spain. The specifications of the amplifiers have been elaborated by the NIKA2 group. In total, NIKA2 is composed of about 2,900 pixels and is equipped with twenty feed-lines. Thus, it employs twenty cryogenics amplifiers (four for the 150\,GHz array and eight for each of the 260\,GHz arrays). The polarisation of the LNA stages is provided by a custom electronics box remotely controlled and allowing the optimisations of the biases according to the slightly different characteristics of the front-end High Electron Mobility Transistors (HEMT). 

\begin{figure}
\begin{center}
\includegraphics[angle=0,width=0.45\textwidth]{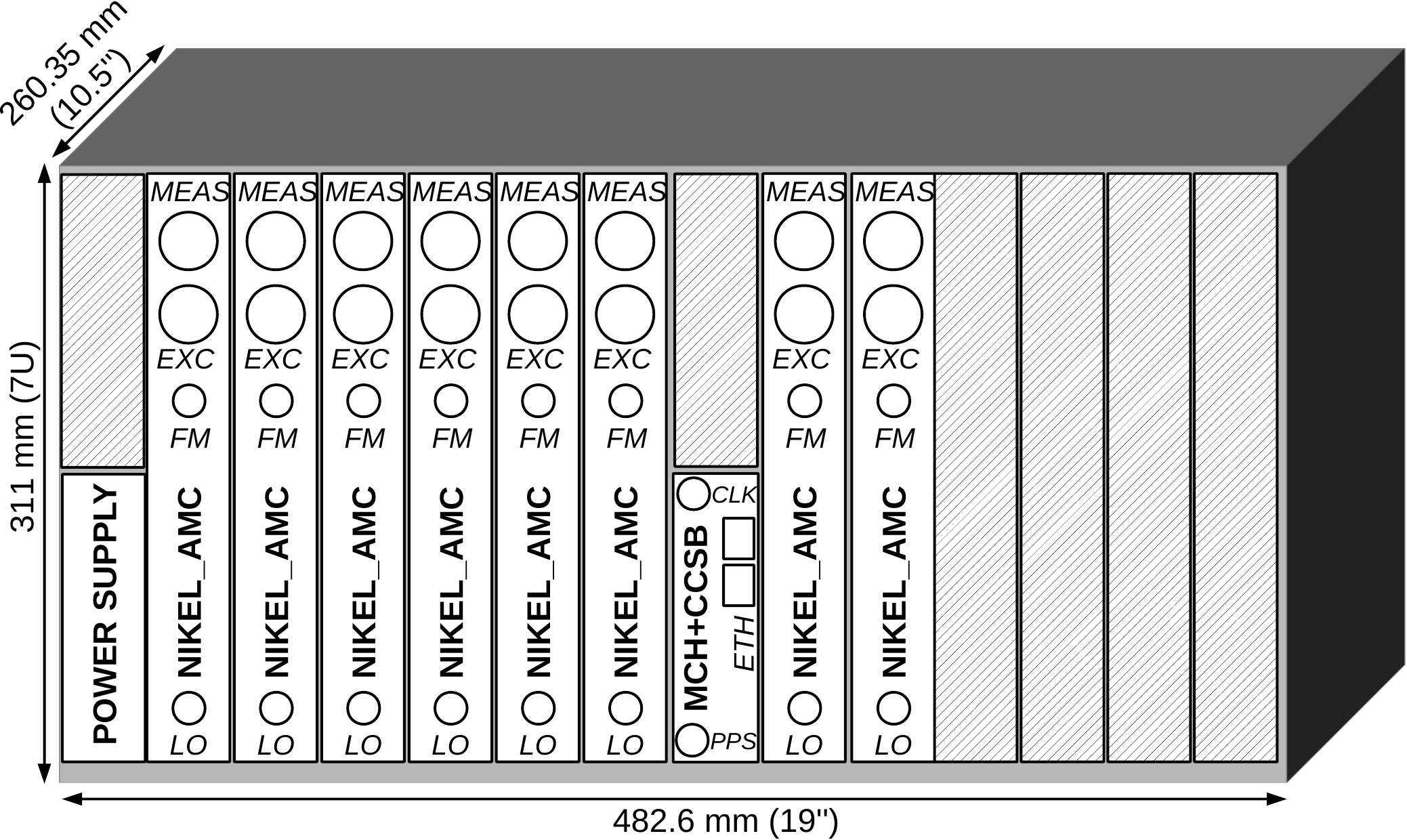}
\caption{Overview of one array readout electronics crate.
It is equipped with 4 (or 8) readout boards lodged in advanced mezzanine card slots (NIKEL\_AMC), one central and clocking and synchronisation board (CCSB) mounted on the MicroTCA Carrier Hub (MCH) and one 600\,W power supply.
The crate allocated to the 150\,GHz channel uses four NIKEL\_AMC boards while the others use eight NIKEL\_AMC boards. The actual power consumption is around 500\,W per crate for eight boards and 300\,W for four boards. The weight of each crate is around 12 kg.
\label{crateFig}}
\end{center}
\end{figure}

The warm electronics required to digitise and process the 2,900 pixels' signals were specifically designed for that purpose;
it is composed of twenty readout cards (one per feed-line) named New Iram Kid ELectronic in Advanced Mezzanine Aard format (NIKEL\_AMC).
As shown in Fig.~\ref{crateFig}, the cards are distributed in three micro-Telecommunication Computing Architecture (MTCA) crates.
A central module, composed of a commercially available Mezzanine Control Hub (MCH) and of custom-made mezzanine boards, is used to distribute a 10\,MHz rubidium reference clock (CLK) and a pulse per second (PPS) signal provided by a global positioning system (GPS) receiver and to control the crate. The synchronisation between the boards and the different crates is ensured by these common reference signals. The electronics is fully described in previous papers (\cite{Bourrion2012,Bourrion2016}).

In summary, the NIKEL\_AMC is composed of two parts: the radio-frequency (RF) part and the digitisation and processing section.
The integrated RF part ensures the transition from and to the baseband part.
It uses the local oscillator (LO) input to perform up and down conversions.
To instrument the 150\,GHz array (resonances from 0.9\,GHz to 1.4\,GHz) and the 260\,GHz arrays (resonances from 1.9\,GHz to 2.4\,GHz), the used LO input frequencies are 0.9\,GHz and 1.9\,GHz,  respectively. The digitisation and signal processing, which is done at baseband, relies on channelised digital down conversion (DDC) and their associated digital sine and cosine signal generators and processors.
The processing heavily relies on field programmable gate arrays (FPGA) while the interfacing to and from the analog domain is achieved by 1\,GSPS analog to digital and digital to analog converters (ADC and DAC, respectively).
The electronics covers a bandwidth of 500\,MHz and handles up to 400\,KIDs in this bandwidth. In NIKA2 about 150 KIDs per board are instrumented, leaving room for placing a number of dark and/or off-resonance excitation tones, and allowing for future developments of the instrument. 
It must be noted that for implementation reasons (\cite{Bourrion2012,Bourrion2016}) the excitation signal, nominally covering 500\,MHz, is constructed by five DACs, each spanning 100\,MHz.

\subsection{Laboratory tests}
\label{Laboratory tests}

NIKA2 has been pre-characterised in the laboratory under realistic conditions. In order to deal with the absence of the telescope optics, we have added a corrective lens at the cryostat input window. This lens generates an image of the focal planes onto our "sky simulator", as described in previous papers (\cite{Catalano2014}, \cite{Monfardini2011}). A sub-beam-sized, that is, point-like, warm source, moved in front of the sky simulator by means of an x-y stage, allows beams and array geometry (e.g. pixel-per-pixel pointing) characterisation. The sensitivity is calculated by executing calibrated temperature sweeps of the sky simulator, and measuring the signal-to-noise ratio. A photometric model has been developed based on ray-tracing simulations. The overall transmission of the instrument, mainly determined by the lenses and the filters, lies around 35\%. The filters have been individually characterised in Cardiff, while the lens transmission has been measured at IRAM, Grenoble. On top of that, the quantum efficiency of the detectors, integrated in the band of interest and calculated from ab-initio electromagnetic simulations, is between 60\% (260\,GHz arrays) and 80\% (150\,GHz array). The three-dimensional (3D) electromagnetic simulations, for the Hilbert design adopted in NIKA2, are confirmed by millimetre-wave vector analyser measurements (\cite{Roesch2012}).

\begin{figure}[h]
\begin{center}
   \centering
    \includegraphics[width=1.0\linewidth]{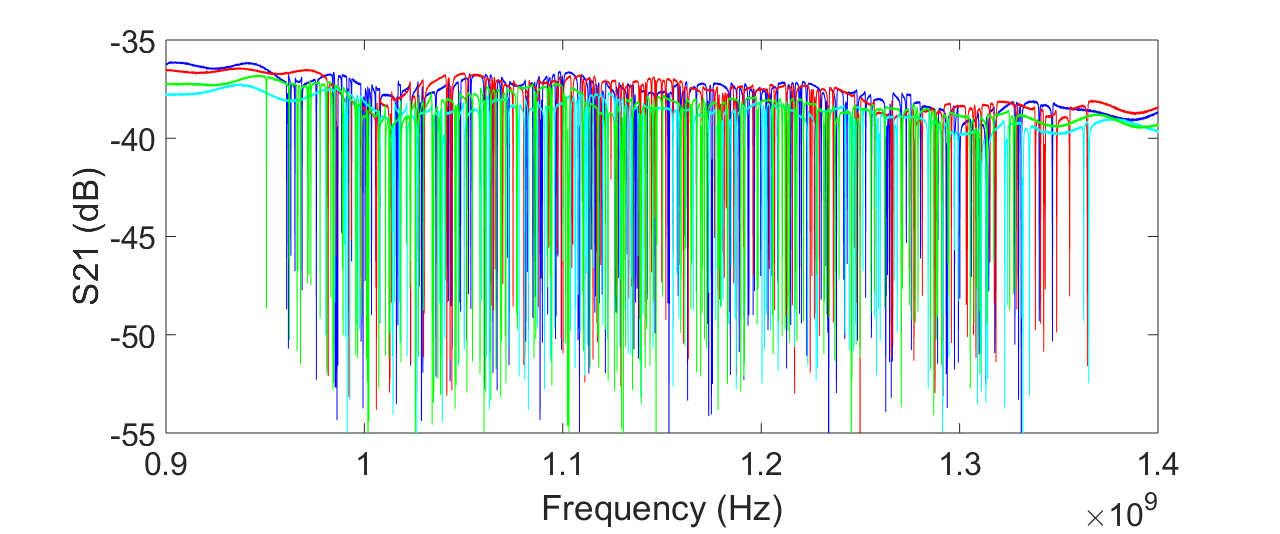}
    \caption{(Colour online) Resonance sweep for the four feedlines of the 150\,GHz array. The lines 1,2,3,4 are  shown in blue, red, cyan and green, respectively. The y-axis represents the transmission of the feedline (parameter S21) and is expressed in dB. Each dip corresponds to a resonance/pixel. At least 94\% of the 616 pixels are identified with a resonance and are thus sensitive to incoming radiation.}
         \label{VNA}
\end{center}
\end{figure}

The frequency sweep of the four lines connected to the 150\,GHz array is shown in Fig. \ref{VNA}. The number of identified resonances over the twenty feedlines exceeds 90\% when compared to the number of pixels implemented by design. 

\begin{figure}[h]
\begin{center}
   \centering
    \includegraphics[width=1.0\linewidth]{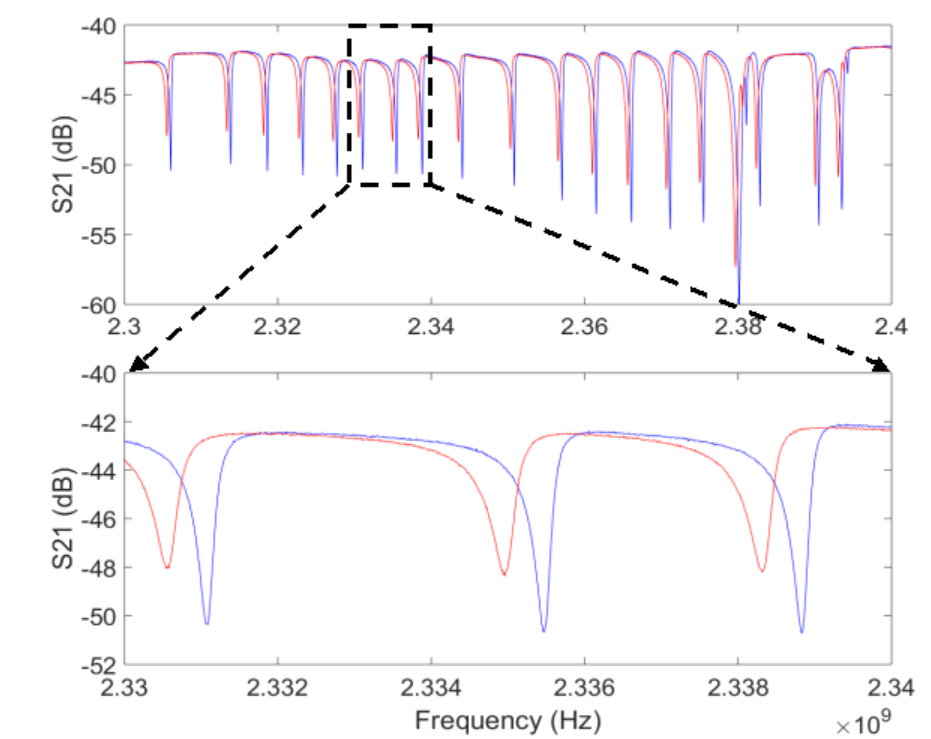}
    \caption{(Colour online) Responsivity estimation using the sky simulator. \underline{Top panel:} frequency sweep of a portion of one particular feedline operating at 260\,GHz. \underline{Bottom panel:} zoom on three typical resonances. In both panels we plot the S21 transmission parameter (dB) against the frequency. Blue lines: cold sky simulator ($T_{SS} \sim 80\,K$). Red lines: 300\,K background. The measured average responsivity, that is, the shift in frequency per unit temperature background variation, is around 2\,kHz/K for the 260\,GHz arrays and 1\,kHz/K in the case of the 150\,GHz array.}
         \label{Shift_f}
\end{center}
\end{figure}

The measurable quantity, proportional to the incoming power per pixel, is the shift in frequency of each resonance (pixel) (\cite{Swenson2010}). This is the reason why our noise spectral densities are expressed in Hz/Hz$^{0.5}$, and the \textit{Rayleigh-Jeans} responsivities are given in kHz/K. By sweeping the temperature of the sky simulator, we have estimated average responsivities around 1 and 2\,kHz/K at 150 and 260\,GHz,  respectively. The array-averaged frequency noise levels for the two bands are about 1 and 3\,Hz/Hz$^{0.5}$, resulting in noise equivalent temperature (NET) of the order of 1 and 1.5\,mK/Hz$^{0.5}$ per pixel at 150 and 260\,GHz, respectively. These figures are calculated at a representative sampling frequency of 5\,Hz. An example measurement based on the sky simulator, and determination of the responsivity, is reported in Fig.~\ref{Shift_f}. 

Noise spectra recorded in the laboratory have been fully confirmed with NIKA2 when operating at the telescope. Please refer to Section~\ref{Noise and sensitivity} for a more detailed discussion concerning the noise properties. 

\begin{figure}[h]
   \centering
    \includegraphics[width=1.0\linewidth]{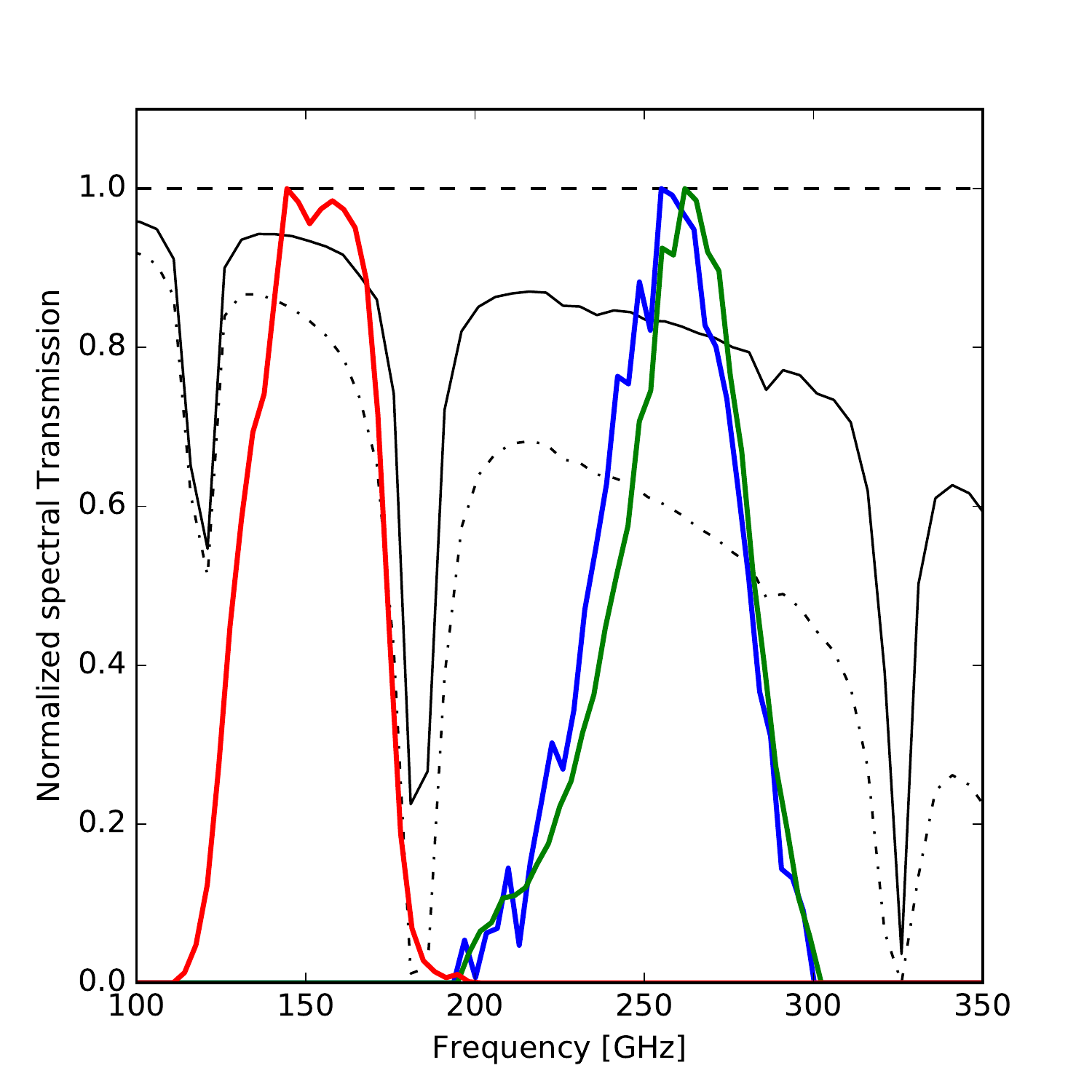}
      \caption{(Colour online) NIKA2 spectral characterisation for the two 260 GHz arrays, H (A1, blue) and V (A3, green) measured in the NIKA2 cryostat, and for the 150\,GHz array (A2, red) measured in a test cryostat equipped with exact copies of the NIKA2 band-defining filters. The band transmissions are not corrected for Rayleigh-Jeans spectrum of the input source. We also show for comparison the atmospheric transmission (\cite{Pardo2002}) assuming 2\,mm of precipitable water vapour (PWV), that is, very good conditions, and 6\,mm PWV, that is, average conditions.
         \label{Fig4}}
\end{figure}

The spectral characterisation of the arrays and the overall optical chain of NIKA2 (Fig.~\ref{Fig4}) has been achieved in the Grenoble laboratory using a Martin-Puplett Interferometer (MpI) built in-house (\cite{Durand2008}) and specifically dedicated to characterisation.
The band transmission in Fig.~\ref{Fig4} was measured using a Rayleigh-Jeans spectrum source.
The two arrays operating at 260\,GHz, mapping different polarisations, exhibit a slightly different spectral behaviour, probably due to a tiny difference in the silicon wafer and/or Aluminium film thicknesses. The observed shift of the peak frequency, 265\,GHz for the V (A1) array versus 258\,GHz for the H (A3), can be explained by an approximately 5-micron change in the substrate thickness. The so-called 1\,mm atmospheric window is not completely filled. This was designed for the first generation of detectors in order to ensure robustness against average atmospheric conditions and to optimise the overall observing efficiency. A possible future upgrade of NIKA2, oriented towards even better sensitivity in very good atmospheric conditions, would be straightforward. 

The sky simulator enabled also a rough but crucial estimation of the parasitic radiation. By comparing measurements obtained at several sky simulator distances with respect to the cryostat window, we determined an equivalent 15\,K additional focal plane background due to the ambient temperature stray radiation. This is lower than the very best equivalent sky temperature at Pico Veleta ($\approx 20\,\textrm{K}$), and confirms that NIKA2 is not significantly affected by this effect. In comparison, in NIKA we had estimated around 35\,K additional background, slightly limiting the performance. 
In summary, the overall performance of the instrument, measured preliminarily in the laboratory, is in line with the NIKA2 specifications, paving the way for the installation at the telescope described briefly in the following Section.

\subsection{The integration at the telescope}
\label{The integration at the telescope}

\begin{figure}[h]
   \centering
    \includegraphics[width=.85\linewidth]{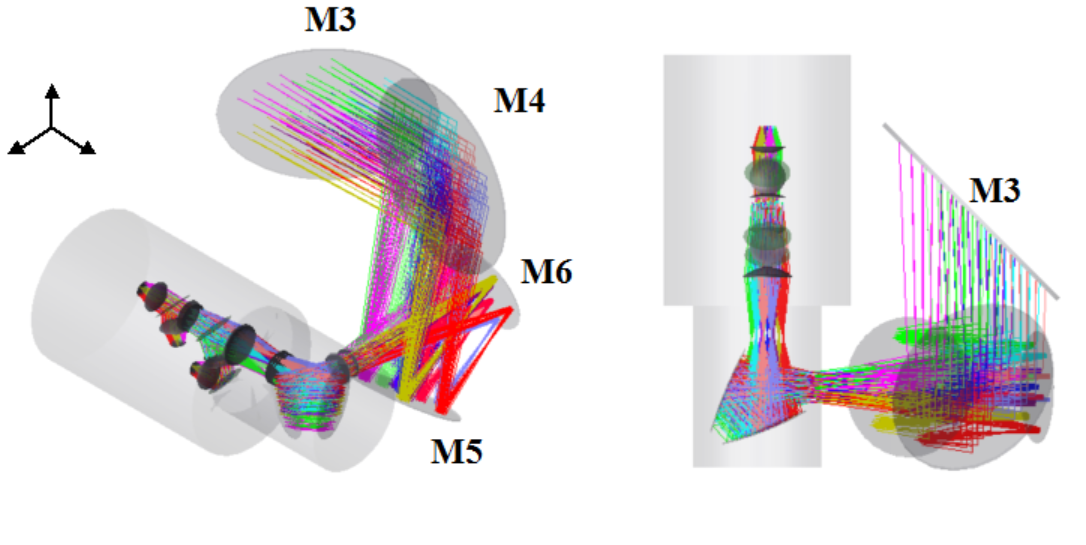}
      \caption{(Colour online) \underline{Left:} Isometric view of the cabin optics scheme, illustrating the mirrors M3, M4, M5 and M6. The ideal case in which the elevation angle is zero degrees is shown. \underline{Right:} Top view of the cabin optics feeding NIKA2.}
         \label{figCabin}
\end{figure}

NIKA2 was transported from the Grenoble integration hall to the observatory at the end of September, 2015. Successful installation of the instrument took place in early October, 2015, at the IRAM 30-meter telescope on Pico Veleta (Sierra Nevada, Spain). To prepare this installation, the optics of the receiver cabin (M3, M4, M5 and M6) had been modified in order to increase the telescope field-of-view up to the 6.5\,arcminutes covered by NIKA2. M3 is the Nasmyth mirror attached to the telescope elevation axis. M4 is a flat mirror that can be turned manually in order to feed the beam either to NIKA2 or to heterodyne spectroscopic instruments (\cite{Carter2012}, \cite{Schuster2004}). The M5 and M6 curved mirrors are dedicated to the NIKA2 camera. The configuration of the optics in the cabin, for an elevation $\delta = 0~\textrm{degrees}$, is drawn in Fig.~\ref{figCabin}. Not shown nor discussed, M1 and M2 are the telescope primary mirror and its sub-reflector,  respectively. 

The whole installation, including the cabling of the instrument, was completed in about three days. The pulse-tube pipes, which are 60 metres long, run through a derotator stage in order to connect the heads in the receiver cabin (rotating in azimuth) and the compressors located in the telescope basement (fixed). A single 1 Giga-bit ethernet cable ensures the communication to and from the NIKA2 instrument. The forty radio-frequency connections (twenty excitation lines, twenty readouts) between the NIKEL\_AMC electronics and the cryostat, located on opposite sides of the receivers cabin, are realised using 10-meter-long coaxial cables exhibiting around 2~dB signal loss at 2~GHz. This is acceptable, considering that the signal is pre-amplified by about 30~dB by the LNAs. 

The optical alignment between the instrument and the telescope optics has been achieved using two red lasers. The first was set shooting perpendicularly from the centre of the NIKA2 input window, through the telescope optics and reaching the vertex and M2. The second laser was mounted on the telescope elevation axis at the M3 position, reaching then, through the M4, M5 and M6 mirrors, the NIKA2 window. In both cases, we have adjusted the cryostat position and tilt. NIKA2 is equipped with an automatic system of pneumatic actuators and position detectors able to adjust the cryostat height and tilt and to keep it stable down to a few tens of microns precision. 

\begin{figure}[h]
   \centering
    \includegraphics[width=.85\linewidth]{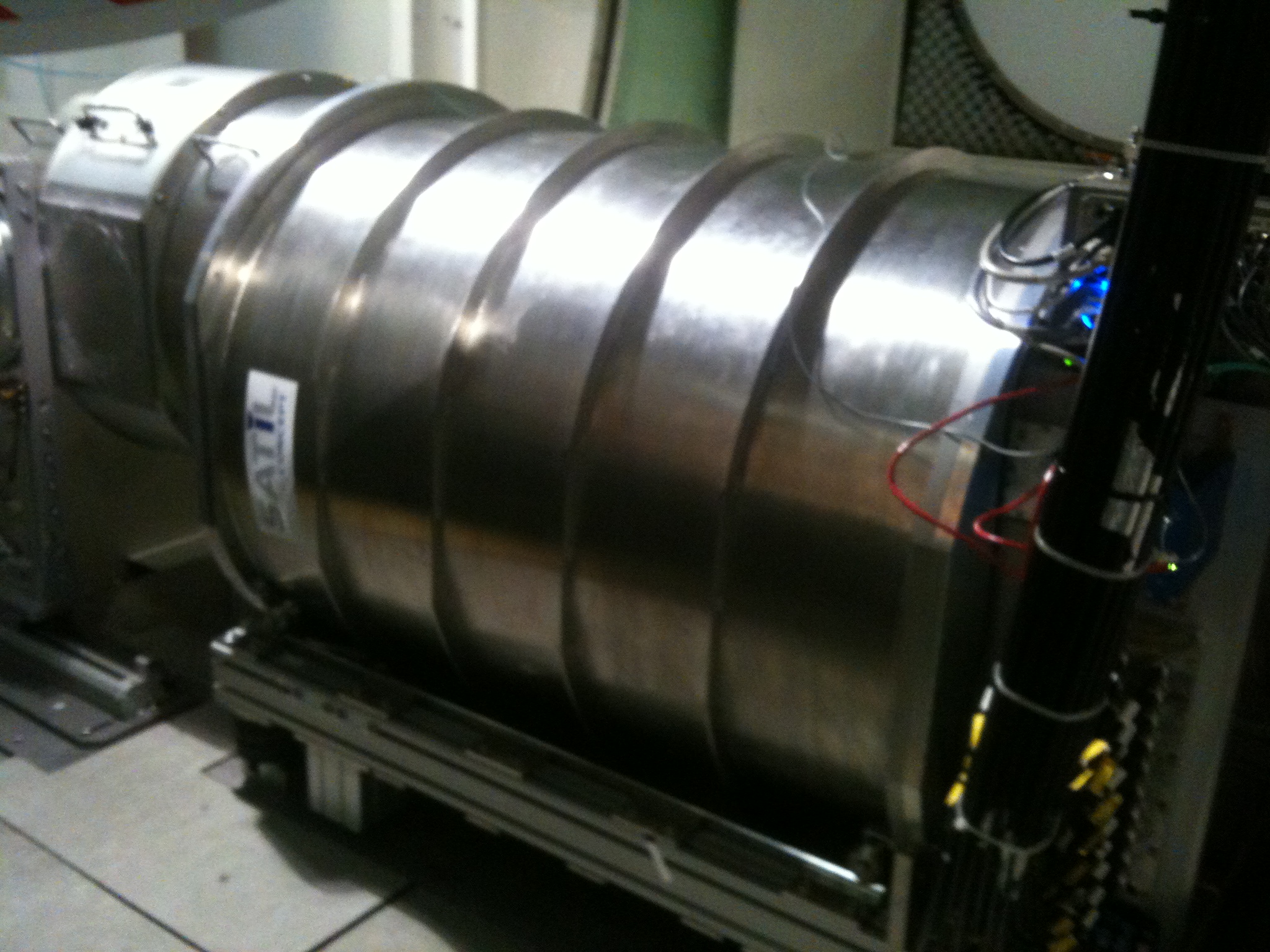}
      \caption{(Colour online) A picture of the NIKA2 cryostat installed in the 30-meter telescope receivers cabin. Picture taken in October, 2015.}
         \label{Fig5}
\end{figure}

The first cryostat cooldown started immediately after the installation, and was achieved after the nominal four days dedicated to pre-cooling, followed by less than 24 hours during which the helium isotopes mixture is condensed in the so-called "mixing chamber". The first-light tests demonstrated that all the detectors were functional and exhibited responsivity and noise in line with the laboratory measurements presented in the previous Section. The preliminary results of the initial technical runs are presented in \cite{catalano2016}.


\section{Measurement principle}
\label{Measurement principle}

At the telescope, the NIKA2 acquisitions on a given source are split into single observational blocks referred to as "scans". In each scan, the source is moved across the field-of-view, typically by constant-elevation pointing sweeps defined as "sub-scans". A particular scan, in which the elevation of the antenna is moved by large amounts, is named "skydip". The skydip allows us to measure the effective atmosphere temperature and calibrate the sky opacity corrections, as explained in Section~\ref{Atmospheric attenuation correction}.

The photometry can be reconstructed down to the required precision thanks to three distinct procedures that are applied during sky observations. First, we implement a real-time electrical calibration acting directly on the KID that is specific to NIKA and NIKA2 (Sect.~\ref{Internal detectors calibration}). Second, KID measurements are dynamically adapted to the sky background (Sect.~\ref{Sky background matching}). Third, the atmosphere opacity correction is calculated in real time thanks to the large dynamic range and linearity of the detectors (Sect.~\ref{Atmospheric attenuation correction}).

\subsection{Internal electrical detector calibration}
\label{Internal detectors calibration}

When radiation is absorbed in a KID, it breaks part of the superconducting carriers (Cooper pairs) and creates a non-thermal excess of unbound electrons (quasi-particles). This changes the impedance of the film and shifts the resonance frequency $f_r$ of the KID to lower values.
The standard way to read a pixel (resonator) is to excite it with a tone at a frequency $f_t$ and monitor how the in-phase ($I$) and in-Quadrature ($Q$) components of the transmitted signal are modified by the changes in its resonance frequency $f_r$. For NIKA2 we adopted the strategy already developed and tested in NIKA. Instead of using an excitation at a fixed frequency $f_t$, we rapidly ($f_{mod} \approx 500\,Hz$) modulate between two different readout tones, $f_t^+$ and $f_t^-$, ideally placed just above and just below $f_r$. The tones are separated by $df=f_t^+-f_t^-$, much smaller than the resonance width. This modulation technique allows us to measure, for every data sample, both the values of $I$ and $Q$ and the variation $dI$, $dQ$ that is induced by the chosen frequency shift $df$. When the optical power on the detectors changes by an amount $\Delta P_{opt}$, a variation $\Delta I$, $\Delta Q$ is observed between successive data samples, which are acquired at a rate of $f_{sampling} = 24\div48\,Hz \ll  f_{mod} $. The $dI$, $dQ$ values can then be used as a calibration factor to associate to the observed $\Delta I$, $\Delta Q$ , the corresponding change in the resonance frequency $\Delta f_r$, and thus measure $\Delta P_{opt}$. A full description of the modulated readout technique is provided in \cite{Calvo2013}.

The advantage of this solution is that the $dI$, $dQ$ values are evaluated for every data sample. If the load on the detectors changes (e.g. due to variations in the atmosphere opacity), the exact shape of the resonance feature of each pixel will change. However, since the calibration factor $dI$, $dQ$ is updated in real time, this effect is taken automatically into account. The photometric accuracy of the instrument, through the KID linearity in terms of $f_r$, is thus strongly improved.

Furthermore, knowing both the $I$, $Q$ and the $dI$, $dQ$ values we can also estimate the difference between $f_r$ and $f_t$. In the ideal situation, these two frequencies should coincide. In reality, changes in the background load can make the resonances drift by a large amount. In such a case, the modulated readout allows us to rapidly readjust $f_t$ to the instantaneous value of $f_r$. This ensures an optimal frequency bias and prevents any degradation in the sensitivity of the detectors. This returning is scheduled, as discussed in more detail in Sect.~\ref{Sky background matching}, between different scans or sub-scans and does not affect the data taken during the proper integration.

\subsection{Sky background matching}
\label{Sky background matching}

During ground-based observations, the radiation load per pixel is determined by the atmospheric transmission and pointing elevation. The load itself is variable in time due to the atmosphere opacity fluctuations around its mean value. The KID tone-frequency load-matching procedure, which we call "tuning", is performed in a specifically dedicated sub-scan at the beginning of each scan and in the lapse of time between two subsequent scans. The tuning procedure is usually performed as a two-step process. First, a common shift is applied to all KIDs in order to match the instantaneous average sky background. Second, the KIDs are individually adjusted by fine-tuning their position. The two steps, depending on the weather conditions, can be executed separately. The versatility of the tuning procedure allows us to keep track of the KID resonance positions even under variable observing conditions, or when the elevation is changed strongly, for example during skydips. The complete tuning, including a verification of the correct frequencies adjustment, is completed in less than 2 seconds. 

The tuning procedure requires real-time synchronisation of the NIKA2 camera with the telescope control system. This is achieved by directly receiving and interpreting the telescope status messages. These messages are broadcast by the telescope server over the NIKA2 private network at a rate of 8\,Hz. The interpreted messages (e.g. begin and end of scans and sub-scans) are recorded in the NIKA2 raw data files. In addition, off-line accurate ($< 0.1~ \textrm{msec}$) synchronisation of the telescope attitude file and the NIKA2 raw data is obtained by monitoring the PPS (Pulse Per Second) signal. This signal, as mentioned in Sect.~\ref{The readout electronics}, is generated by the telescope control system and shared by all the instruments.

\subsection{Atmospheric attenuation correction}
\label{Atmospheric attenuation correction}

The sky maps have to be corrected for the atmospheric absorption. The corrected brightness $S^{corrected}$ is:
\begin{equation}\label{eq:opa}
S^{corrected} =  S^{ground} \cdot e^{ x \cdot \tau_{scan}}
,\end{equation}
where $\tau_{scan}$ is the zenith opacity, for each band, of the atmosphere during the
observation, and $x$ represents the airmass\footnote{By assuming a
  homogeneous plane-parallel atmosphere, the relation between the
  airmass and the elevation of the telescope is taken as $x =
  csc(\delta)=1/sin(\delta)$, where $\delta$ is the average
  elevation.} at the considered elevation.

\begin{figure}
\includegraphics[scale=0.55]{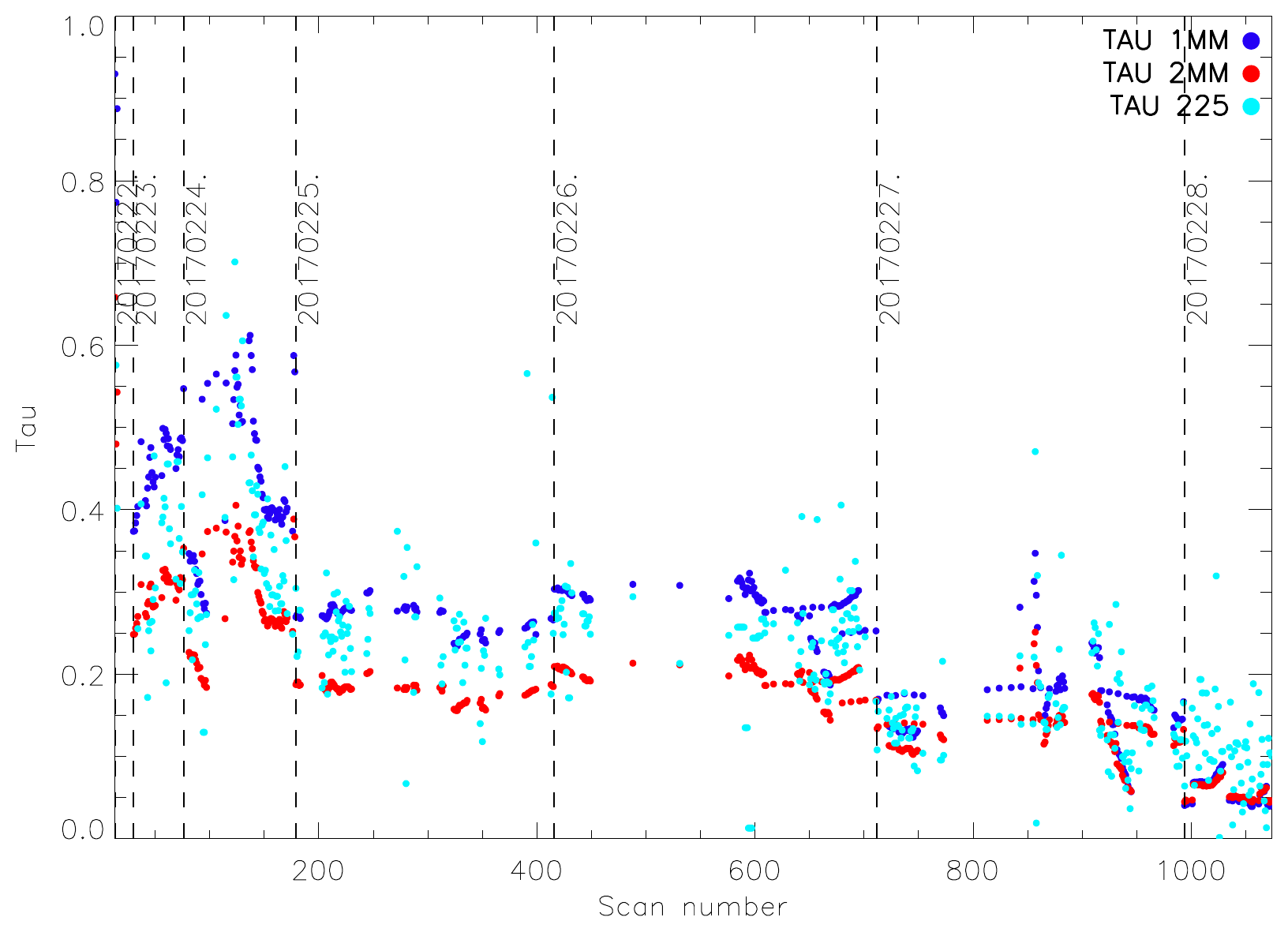}
\caption{(Colour online) Atmospheric opacity as measured from the IRAM 225\,GHz tau-meter (cyan), and from the NIKA2 data at 150 (red) and 260\,GHz (blue) during the February 2017 NIKA2 commissioning campaign. We stress the fact that the IRAM 225\,GHz tau-meter data is not used for the atmospheric correction and is plotted here just for comparison.
  \label{fig:taumeas}}
\end{figure}

NIKA2 has the ability to compute the opacity directly along the line of sight, integrated in NIKA2's exact bandpasses and independently of the IRAM tau-meter operating at 225~GHz. The procedure was successfully tested with NIKA and is described in detail in \cite{Catalano2014}. Indeed, the resonance frequency of each detector is related to the airmass and the opacity as in the following:
\begin{equation}\label{eq:skydip}
f^{i}_{r} = c^{i}_0 - c^{i}_1 \cdot T_{atm}[1 - e^{- x \cdot \tau}],  i = 1 ... N
\label{eq:tau}
,\end{equation}
where $f^{i}_{r}$ is the measurable absolute value of the $i_{th}$ detector resonance frequency (Sect.~\ref{Internal detectors calibration}), $c^{i}_0$ is a constant equal to the resonance frequency at zero opacity, $c^{i}_1$ is the calibration conversion factor in $\mathrm{kHz/K,}$  with $T_{atm}$ we refer to the equivalent temperature of the atmosphere (taken as a constant at $270\,\mathrm{K}$), $\tau$ is the sky zenith opacity, as a function of time, and N is the number of useful detectors in the considered array. 

The coefficients $c^{i}_0$ and $c^{i}_1$ are expected to be constant in time within at least one cooldown cycle; once they are known, Eq.~(\ref{eq:tau}) can be inverted to determine $\tau$ at the corresponding time. The determination of the constants $c^{i}_0$ and $c^{i}_1$ is achieved via a specific scan, nicknamed \emph{skydip}. For such a scan, the telescope performs eleven elevation steps in the range $\delta = 19\div65\,\mathrm{deg}$, regularly spaced in airmass. For each step, we acquire about twenty seconds of time traces to reduce the error in the determination of $f^{i}_{r}$. Several skydips under different weather conditions (hence different $\tau$) are solved simultaneously in order to break the natural degeneracy between the opacity and the responsivity.

We observe that the skydip-fitted $\tau$ values are, as expected, common
between different detectors of the same array. By comparing the results of
different skydips, we have verified experimentally that the coefficients
$c^{i}_0$, $c^{i}_1$ are stable, within the fit errors, on very long time
scales within a cooldown cycle. The coefficients can thus be applied to the
whole observing campaign in order to recover the opacity of each scan. In
Fig.~\ref{fig:taumeas} we present the evolution of the NIKA2 in-band opacities
for several scans of the commissioning run held in February 2017. These are
compared to the IRAM tau-meter readings. We observe a global trend agreement
between the IRAM-tau-meter-suggested opacity (225~GHz) and the NIKA2
values. These latter values show, however, a smaller dispersion. We find an
average ratio between the 150~GHz and the 260~GHz NIKA2-derived opacities of
about 0.6, which is only broadly consistent with model expectations. We notice however that the 150~GHz-to-260~GHz opacity ratio varies significantly for opacities (at 150~GHz) below $0.2$. This effect is likely to be linked to an O$_2$ atmospheric line which becomes saturated. This point is, however, still under investigation.

\section{Observations and performance}
\label{Observations and performance}

The first NIKA2 astronomical light was achieved in October 2015. A first technical run immediately followed. A number of commissioning runs were then carried out between November, 2015, and April, 2017. The commissioning observations were carried out periodically and did not interfere with the heterodyne routine observations scheduled at the telescope. 
Since September, 2016, the instrumental configuration has been fully stable after replacing the 150~GHz array, the dichroic, and the smooth lenses with anti-reflection ones.
In this paragraph we summarise the first results obtained for the characterisation of the instrument performance. More in-depth details concerning the data analysis pipeline and the commissioning results will be given in forthcoming papers (\cite{pipeline} and \cite{commissioning}). The commissioning of the polarisation channel is on-going, and results will also be the object of a future paper. 

We stress that the experience from the use of NIKA2 by external astronomers might lead, in the best case, to further optimisation of the instrument performance. The experience that will be accumulated in the future might eventually allow us to evaluate subtle problems that have not been solved during the commissioning. 

\subsection{Data processing at the telescope}
\label{Data processing at the telescope}

Astrophysical observations are carried out in scans, which typically last for a few minutes up to 20 minutes at most.
The NIKA2 data in intensity are sampled at 23.8418~Hz. In polarimetric mode, the sampling runs at twice this frequency, that is, 47.6836~Hz. For each sample and for each KID, the In-phase (I), the Quadrature (Q) and their derivatives (dI, dQ) of the transfer function of the feed-line and the pixel are recorded. Some extra information like scan number, sub-scan number, and telescope pointing information are also included. \\

Data processing is needed during telescope observations to ensure the scientific quality of the acquired data.
We have developed two sets of tools for real-time and quick-look analysis. The real-time tools run on a dedicated multi-processor acquisition computer and monitor the KID time ordered data and the overall behaviour of the instrument, and decode telescope messages as discussed in Sect.~\ref{Sky background matching}. \\

The quick-look analysis software is run at the end of each scan by the observer to obtain early feedback. The execution time is typically of the order of one minute. It includes both map making capacities and analysis of the produced maps, in terms of photometry, sensitivity and calibration. Indeed, it is used to monitor the pointing of the telescope and its focus (about once every two hours in normal observing conditions) and to give the required instructions to correct for their drifts. This quick-look software has been used extensively during the commissioning. It shares a number of common tools with the off-line processing pipeline that are used to construct the final and optimal sky maps (\cite{pipeline}).

\subsection{Field-of-view reconstruction}
\label{Field of view reconstruction}

The reconstruction of the position of the detectors in the field-of-view (FoV) is mainly based on observations of planets, and in particular Uranus, Neptune, and Mars. We generally perform deep-integration azimuth raster-scan observations at constant elevation. A total of 99 subscans are taken by changing elevation in steps of 4.8$^{\prime \prime}$. The overall footprint of these scans, which we call {\it beam-maps}, is $780^{\prime \prime} \times 470^{\prime \prime}$. We produce a map of the source for each KID with a projected angular resolution of 4$^{\prime \prime}$. These maps are used to derive the KID position on the FoV, the properties of the beam pattern (FWHM and ellipticity) per KID, and the detector inter-calibration. 

\begin{figure}[h]
   \centering
    \includegraphics[trim=2cm 12cm 4.6cm 4.3cm, clip=true,width=0.78\linewidth]{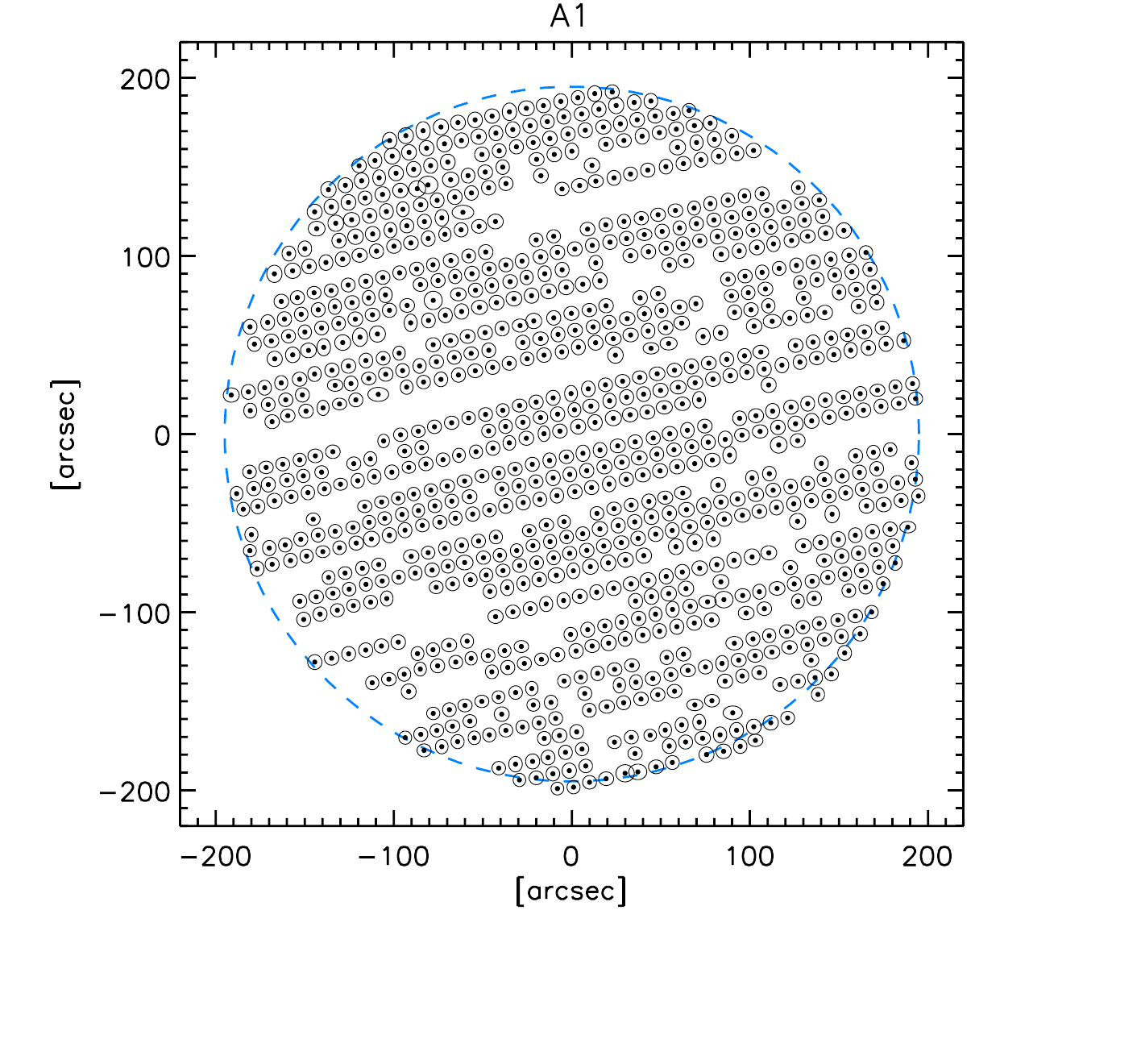}
   \includegraphics[trim=2cm 12cm 4.6cm 4.3cm, clip=true,width=0.78\linewidth]{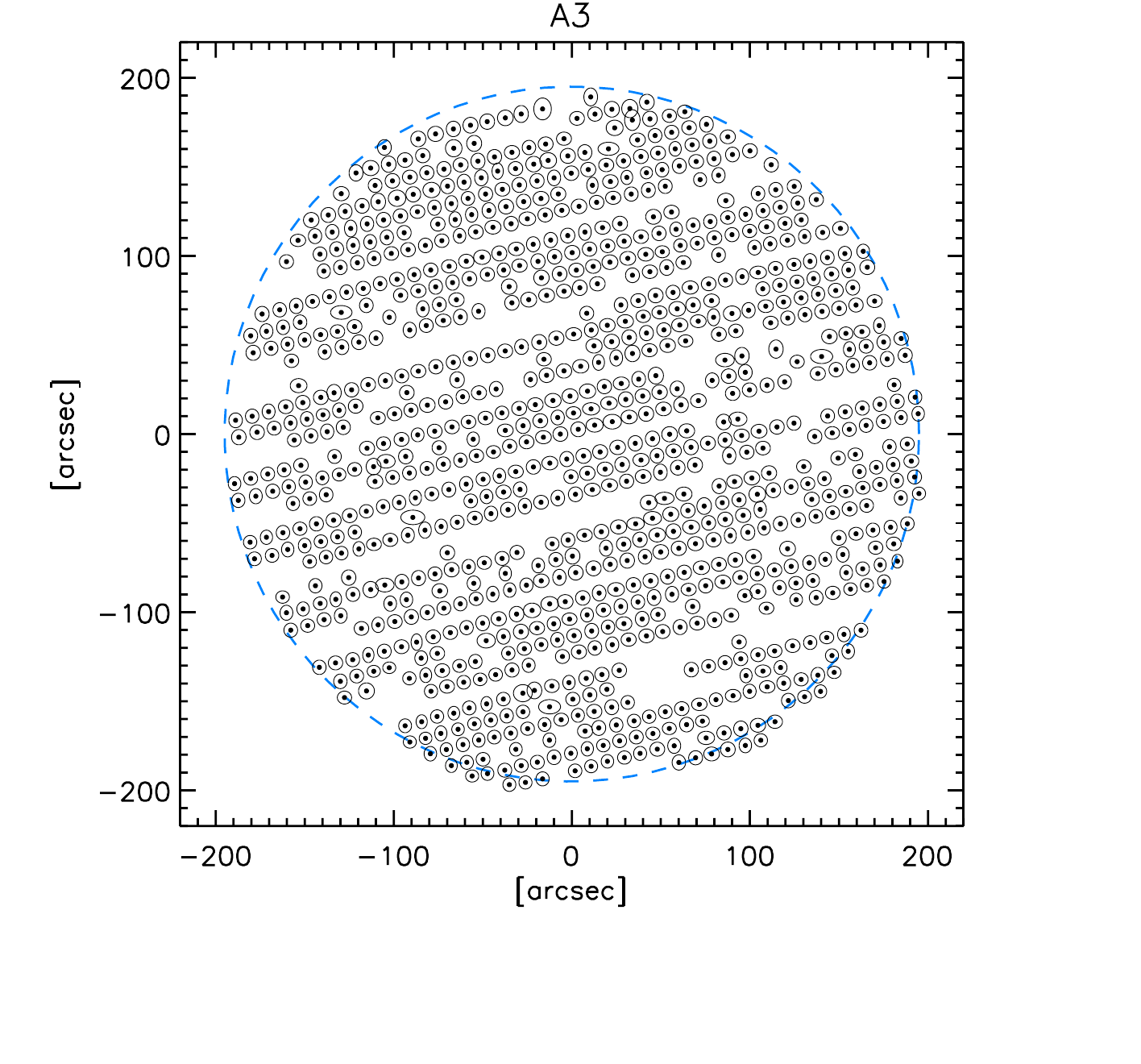}
   \includegraphics[trim=2cm 12cm 4.6cm 4.3cm, clip=true,width=0.78\linewidth]{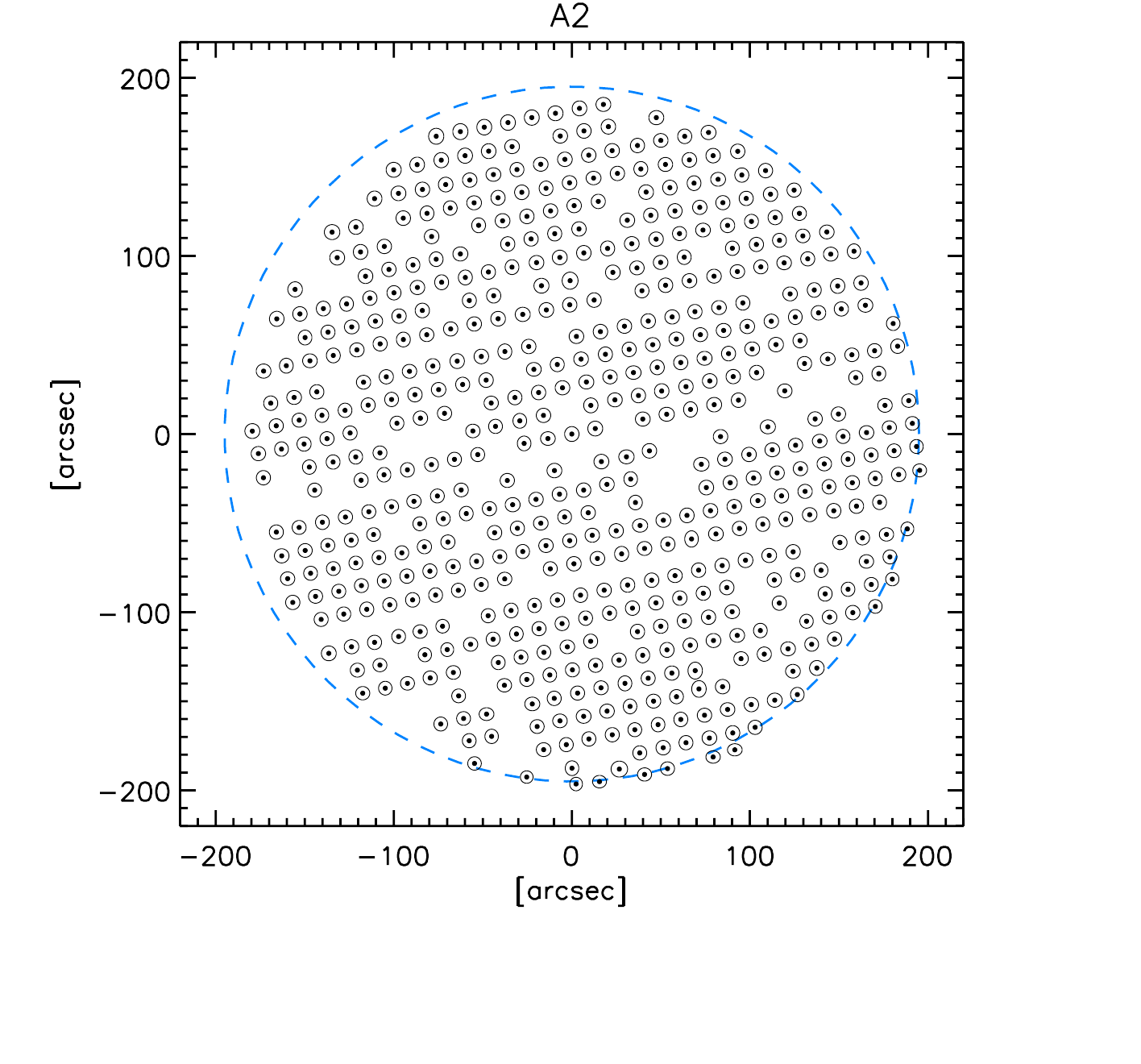}       
      \caption{From top to bottom, detectors positions for arrays A1 (260~GHz-H), A3 (260~GHz-V), and A2 (150~GHz). The three plots show the detectors that have seen the sky and passed the quality criteria for at least two focal plane reconstructions during Run10: 952, 961, and 553 for A1, A3 and A2, respectively. The outer dashed line circle corresponds to the nominal FoV of 6.5~arc-minutes.
         \label{fig:focalplane}}
\end{figure}

\begin{figure}[h]
  \centering
  \includegraphics[clip=true,width=0.68\linewidth]{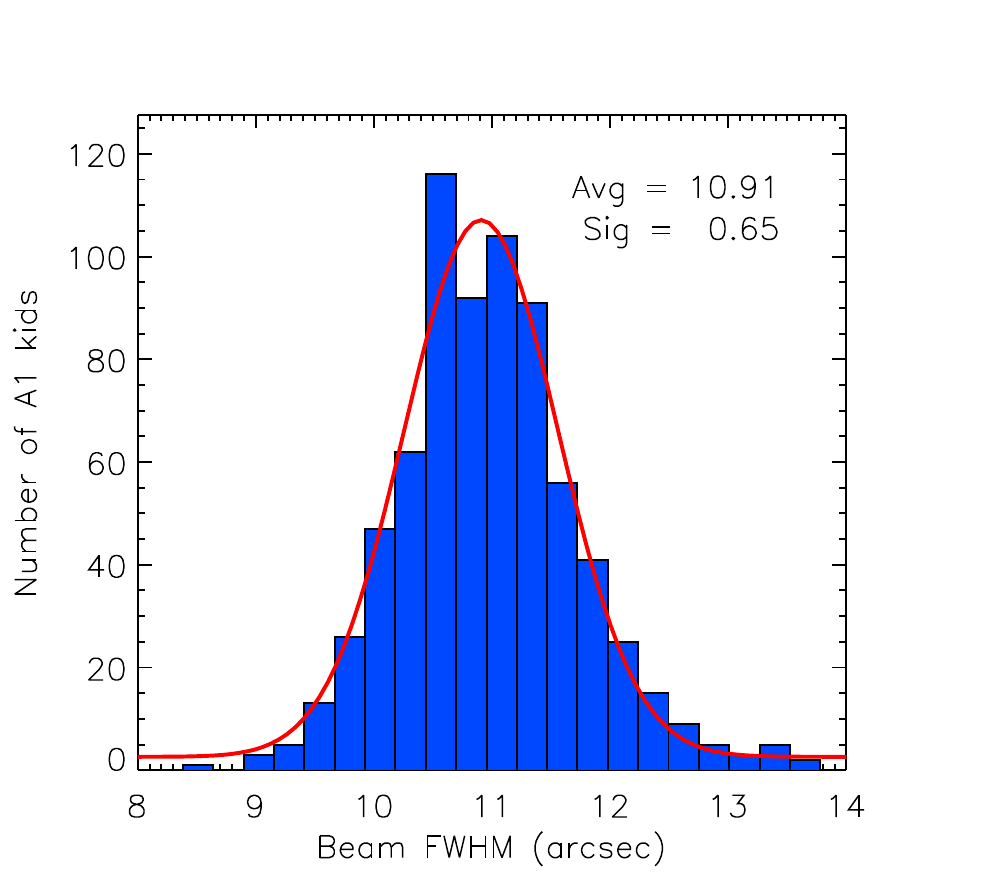}
  \includegraphics[clip=true,width=0.68\linewidth]{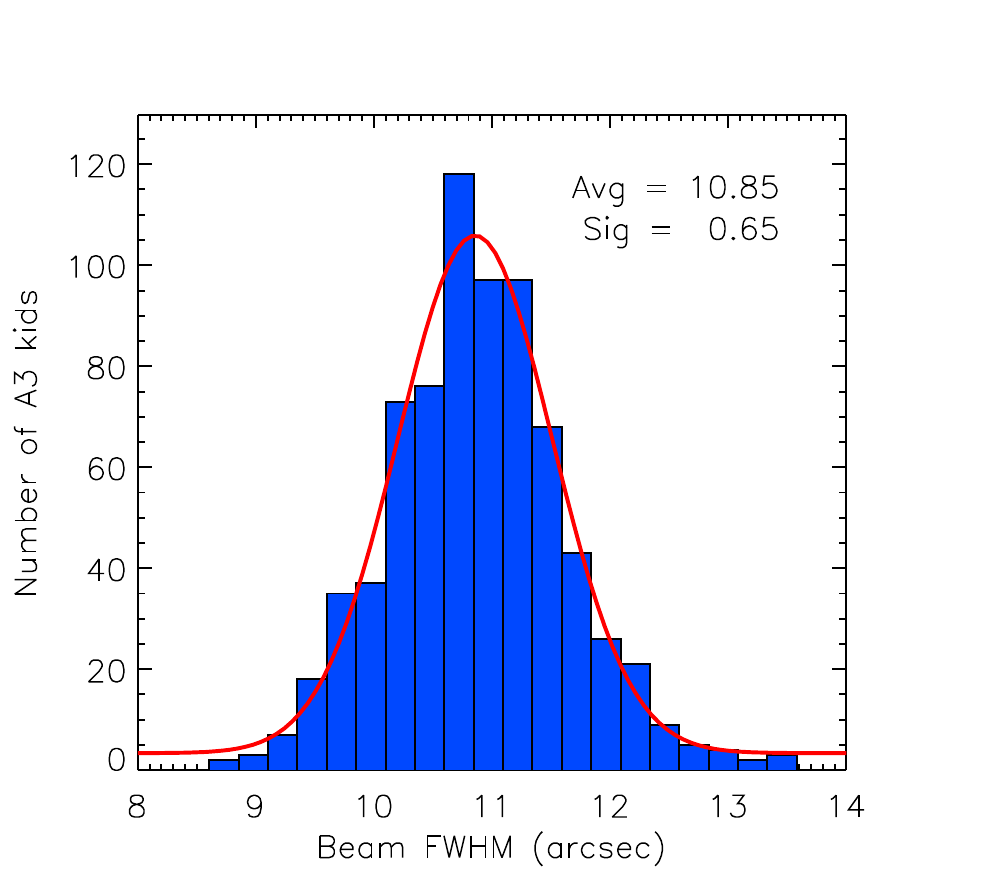}
  \includegraphics[clip=true,width=0.68\linewidth]{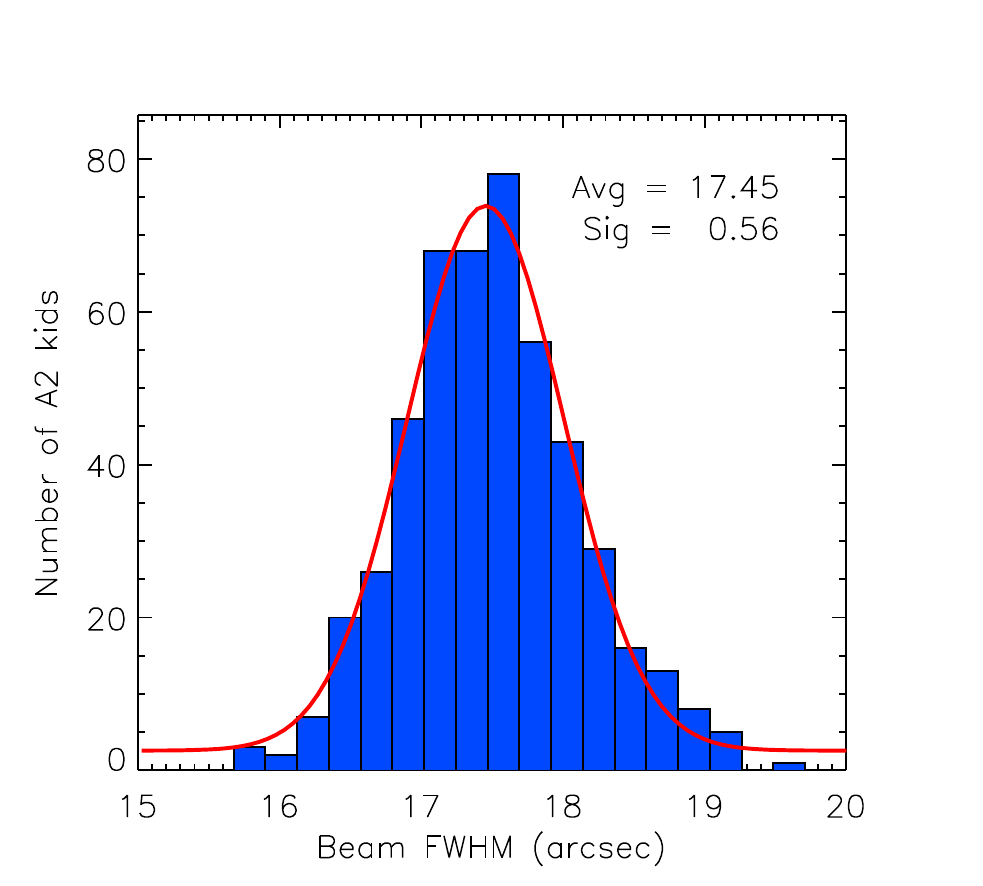}
  
\caption{(Colour online) From top to bottom, main beam FWHM distribution of all valid KID detectors of arrays A1, A3, and A2. The main beam FWHM is the geometrical combination of the two-orthogonal FWHM estimates obtained from an elliptical Gaussian fit on side-lobe masked individual maps per KID (see text). The red curves show a Gaussian fit to the histogram data.}
  \label{fig:focalplane_histo}
\end{figure}

Figure~\ref{fig:focalplane} shows the position of the detectors in the NIKA2 FoV for the two 260~GHz arrays (A1 and A3), and for the 150~GHz array (A2). For each detector the ellipse symbol size and ellipticity are proportional to the main beam FWHM and ellipticity, as derived from a fit to a 2D elliptical Gaussian. To isolate the main beam contribution to the total beam, the sidelobes are masked out using annulus masks centered on the peak signal, of $50\arcsec$ external radius and of internal radius of $9\arcsec$ at $260~\rm{GHz}$ and $14\arcsec$ at $150~\rm{GHz}$. Elliptical 2D Gaussian fits on the masked maps of each detector (individually) provide two orthogonal-direction FWHMs, which are geometrically combined to obtain the main beam FWHM. Figure~\ref{fig:focalplane_histo} shows the distribution of the main beam FWHMs of the arrays A1, A3, and A2 using a beam-map scan of Neptune acquired during the April 2017 commissioning campaign and for average weather conditions. We also show in red the best Gaussian fit to histogram data. We find an average main beam FWHM of $10.9\arcsec$ at $260~\rm{GHz}$ and $17.5\arcsec$ at $150~\rm{GHz}$ in agreement with the main beam estimates from the deep beam map presented in Fig.~\ref{fig:beampattern}. The observed dispersion of about $0.6\arcsec$ is expected from the optics design and its associated field distortions across the FoV of  6.5~arc-minutes. In particular, these distortions lead to an optimal focus, determined by the position of the M2 mirror, that is shifted by 0.2~\rm{mm} between the centre and the outer edge of the FoV.
 
\subsection{Antenna diagram}
\label{Antenna diagram}

\begin{figure}[h]
   \centering
    \includegraphics[width=1.00\linewidth]{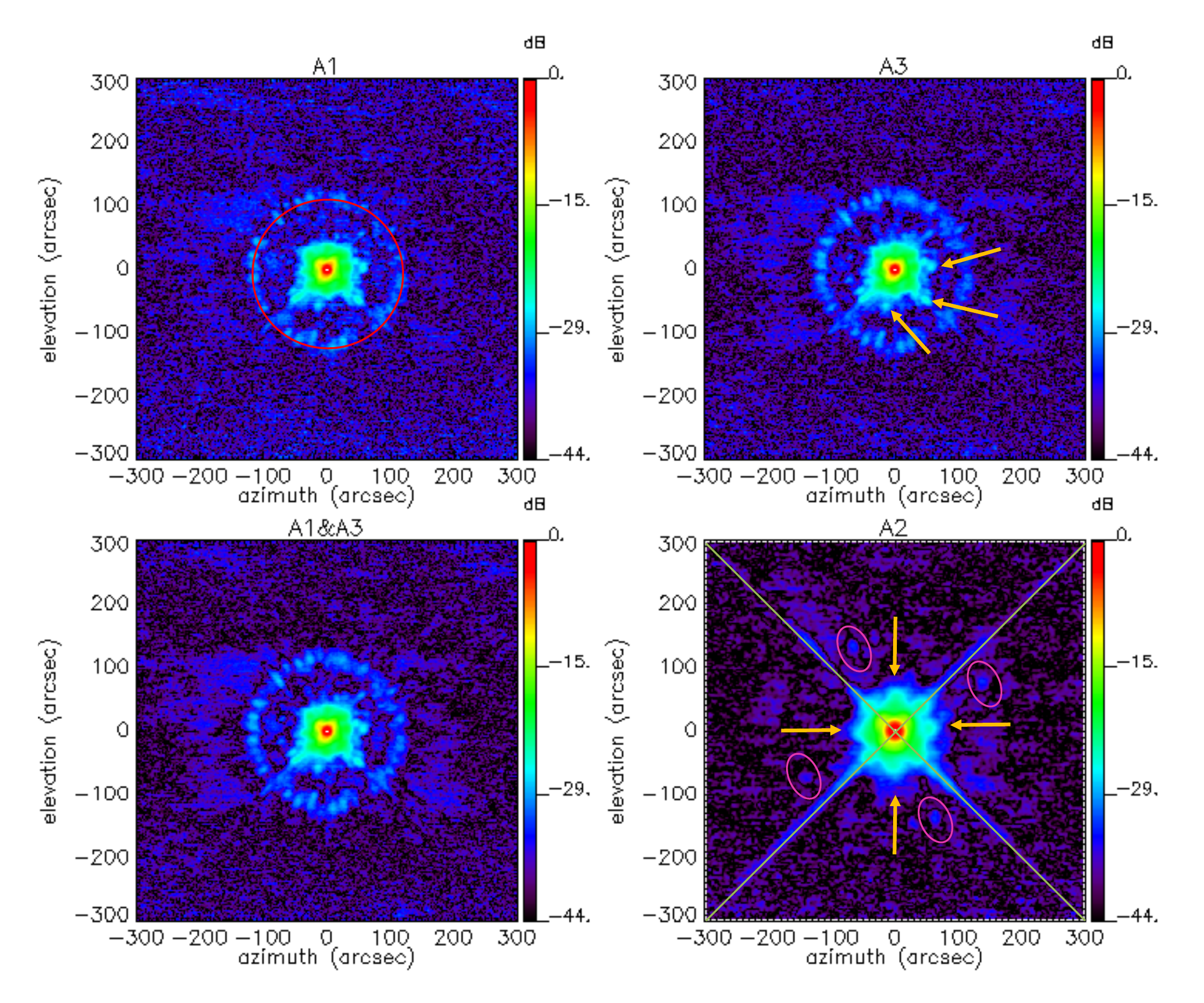}  
      \caption{(Colour online) Measured beam pattern. From upper left to lower right, beam maps of arrays A1 and A3, the combination of the 260 GHz arrays (A1\&A3) and the 150~GHz array (A2) are shown in decibel. These $10^{\prime} \times 10^{\prime}$ maps, have been constructed from the normalised combination of four relatively long scans of bright point sources. Details on the structures present in the maps are given in the text.}
         \label{fig:beampattern}
\end{figure}

To probe the extended beam patterning, we use the same observations as above, and produce a map with all usable detectors. We show in Fig.~\ref{fig:beampattern} the beam pattern as obtained from the optical instrument and telescope response to Uranus for arrays A1, A3, the combination of A1 and A3 (260~GHz), and A2 (150~GHz). The telescope beam is characterised by its main beam, side lobes, and error beams. The main beam is well described by a 2D Gaussian, while the error beams are more complex and have been fitted to the superposition of three Gaussians of increasing FWHM (65$^{\prime \prime}$, 250$^{\prime \prime}$ and 860$^{\prime \prime}$ at 210 GHz) in \cite{greve1998,kramer2013} using observations of the lunar edge with single pixel heterodyne receivers to construct beam profiles.

The maps are consistent with a 2D gaussian main beam of FWHM $11.3^{\prime \prime} \pm 0.2^{\prime \prime}$, $11.2^{\prime \prime} \pm 0.2^{\prime \prime}$, and $17.7^{\prime \prime} \pm 0.1^{\prime \prime}$, for A1, A3 and A2 arrays, respectively. These results are consistent with the mean FWHM of the main beam of individual KIDs shown in Fig.~\ref{fig:focalplane_histo} and are presented in Table~\ref{sumperf}. The colour lines in Fig.~\ref{fig:beampattern} show the relatively complex sidelobes and error beams. In the 260 GHz maps (A1, A3, and combined A1 \& A3) we clearly observe a diffraction ring at a radius of about 100$^{\prime \prime}$ and at -30 dB. The diffraction ring and its spokes are presumably caused by radial and azimuthal panel buckling (\cite{greve1998}). The perpendicular green lines shown in the A2 (150~GHz) map correspond to the diffraction pattern caused by the quadrapod structure supporting M2. In the same map the yellow arrows point to four symmetrical spokes of the error beams. The pink ellipses show spikes in the A2 map. We observe, in the A3 map in Fig.~\ref{fig:beampattern}, some spikes of unknown origin.

Comparing the 2D Gaussian main beam fit to the full beam pattern measurement up to a radius of $250''$, we compute the beam efficiencies defined as the ratio of power between the main beam and this full beam. We find beam efficiencies of $\sim 55$ \% and $\sim 75$ \% for the 260 and 150~GHz channels, respectively.
Heterodyne observations of the lunar edge and of the forward beam efficiency derived from skydips show that a significant fraction of the full beam is received from beyond a radius of 250$^{\prime \prime}$. This fraction is not considered here.

\subsection{On-sky calibration}
\label{On-sky calibration}

The planets Uranus and Neptune were used as the primary calibrators. Their reference flux densities were obtained from the model in \cite{moreno2010, Bendo2013} 
and updated at the mid-date of each session of observations. We use the planet geocentric distance and viewing angle to account for planetary oblateness as provided by the JPL's HORIZONS Ephemeris\footnote{https://ssd.jpl.nasa.gov/horizons.cgi}. By convention, flux densities are given at reference frequencies 150 and 260~GHz for the two channels, respectively. We notice that the reference frequencies are close to the peak frequencies for each channel, which were discussed in Sect.~\ref{Laboratory tests} .

Various observations of Uranus and Neptune with integration times of $\sim 20$ minutes were carried out during each commissioning session resulting in high SNR maps (e.g. Fig. \ref{fig:beampattern}). These observations of strong sources with relatively long integration times were planned to minimise statistical noise in order to determine the level of systematics that characterises the stability of the scale. Their total flux densities were measured from the maps by aperture photometry within a radius of $150''$ where cumulative flux density levelled off smoothly. For this photometry, we used  the solid angle of the beam of the telescope that we could determine for each observation with these strong sources  in using their radial brightness profile computed at the maximum extent $r_{max}=250''$. The beam solid angle  was found to be slightly variable, as expected since no telescope gain dependence on elevation was yet implemented for the processing of the NIKA2 observations, but also because atmospheric conditions might have had an impact. The obtained fluxes were corrected for atmospheric absorption using the atmospheric line-of-sight opacity for the two NIKA2 channels, which was computed as described in \ref{Atmospheric attenuation correction}.

\begin{figure}[h]
\centering
\includegraphics[angle=270,width=1.0\linewidth]{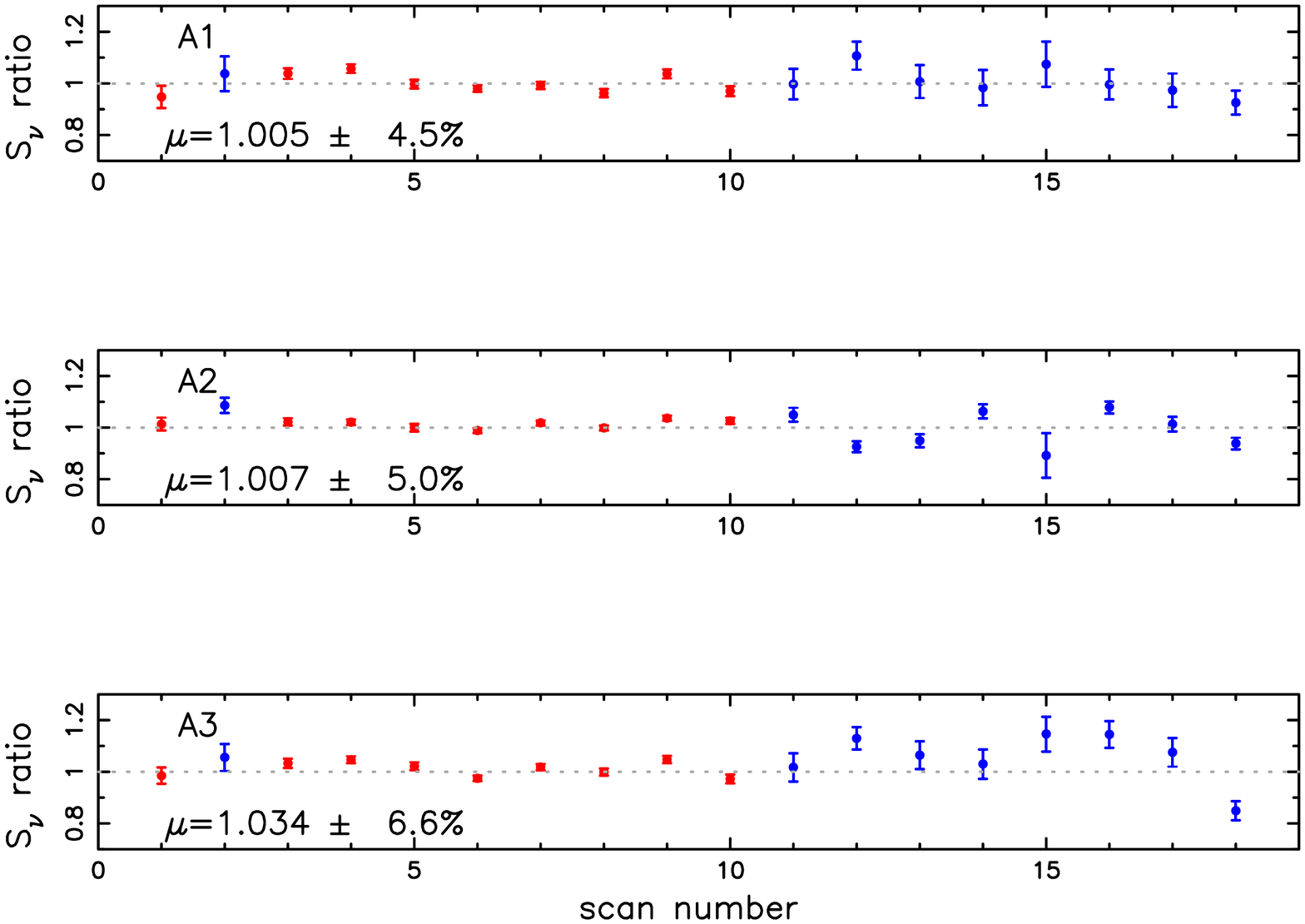}
\caption{(Colour online) Comparison of measured and reference flux densities of the primary calibrators Uranus (red) and Neptune (blue). Their ratios are shown for the three arrays A1 (260~GHz-H), A2 (150~GHz), A3 (260~GHz-V). The mean ratio $\mu$ and relative scatter are provided for each array. The reference flux densities are from \cite{moreno2010,Bendo2013}. The scan numbers are time-ordered: 1 to 10 refer the period 23-28 February, 2017, (fair weather) and 11 to 18 to the  period 19-25 April, 2017, (mediocre weather).}
\label{fig:calibaccuracy}
\end{figure}

The ratios between the fluxes for each individual planet observation and the reference planet flux density for the three NIKA2 arrays are shown in Fig.~\ref{fig:calibaccuracy} for the February and April, 2017, commissioning campaigns.  The mean ratios for the three arrays are close to unity as expected since the planets were used to set the calibration factors in the off-line processing. Overall, the flux density scale is stable at better than $7\%$ for all observations acquired during the two one-week runs (separated by two months) and despite the fact that the instrument was warmed up in between the two sessions. It is noticeable that the scatter around unity in Fig.~\ref{fig:calibaccuracy} is about twice smaller in the first (February) session, which was conducted in significantly better weather conditions.

More precisely, we quantify the stabilities for the February commissioning week via the relative flux rms and find values of 3.6\%, 2.5\% and 2.9\% for arrays A1, A2 and A3, respectively, with atmospheric opacity at 260~GHz between 0.05 and 0.3. Correspondingly, for the April campaign, we find 5.3\%, 6.7\% and 8.6\%  with atmospheric opacity at 260~GHz between 0.25 and 0.6. It is thought that, at the moment, limitations in stability are caused by residual atmospheric fluctuations and uncertainty in opacity corrections.

Nonetheless, the flux density scale of NIKA2 is found to be highly stable and comparable to the level achieved by other modern instruments, such as SCUBA2 (\cite{Dempsey2013}). The other limitation of the scale is absolute calibration that depends on the accuracy of the \cite{moreno2010,Bendo2013} model which is estimated to be 5\% in the millimetre wavelength range. Hence, in combining both limitations, the total uncertainty of calibration with NIKA2 is $10\%$ in mediocre atmospheric conditions and better than $6\%$ in fair conditions ($\tau_{260 GHz} <0.3$). 

\begin{figure}[h]
   \centering
    \includegraphics[trim=1cm 4cm 2cm 12cm, clip=true,width=0.85\linewidth]{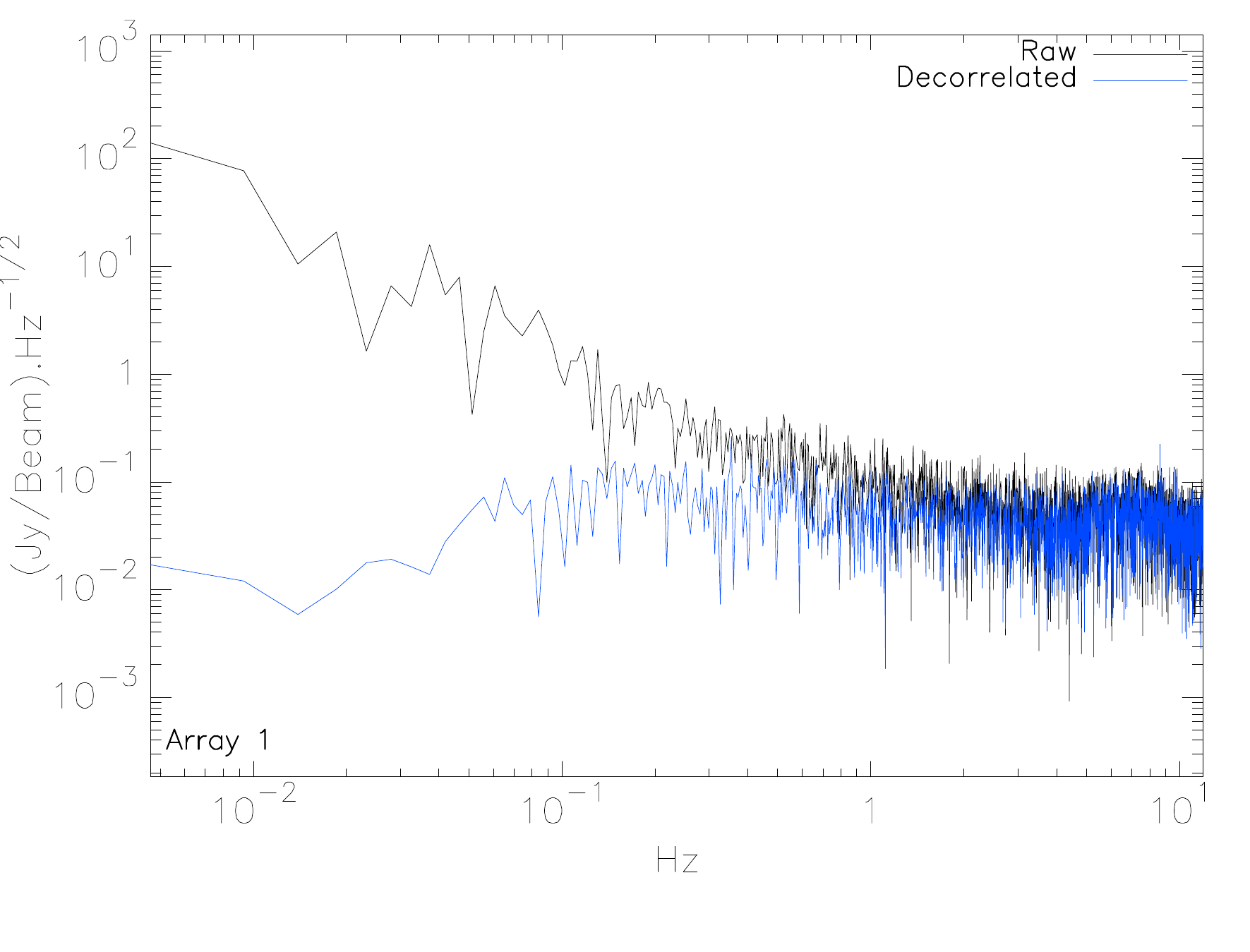}     
       \includegraphics[trim=1cm 4cm 2cm 12cm, clip=true,width=0.85\linewidth]{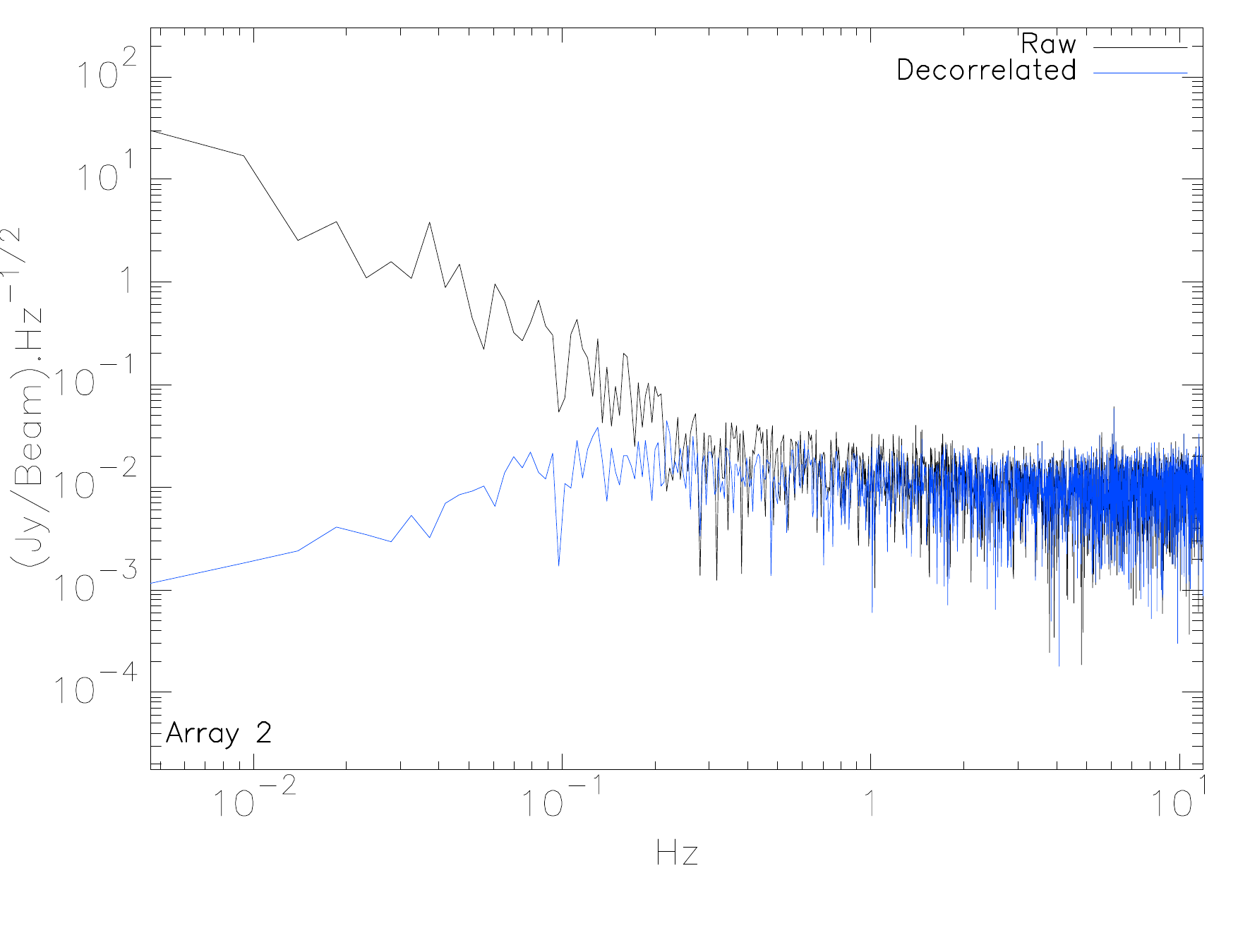}     
   \includegraphics[trim=1cm 4cm 2cm 12cm, clip=true,width=0.85\linewidth]{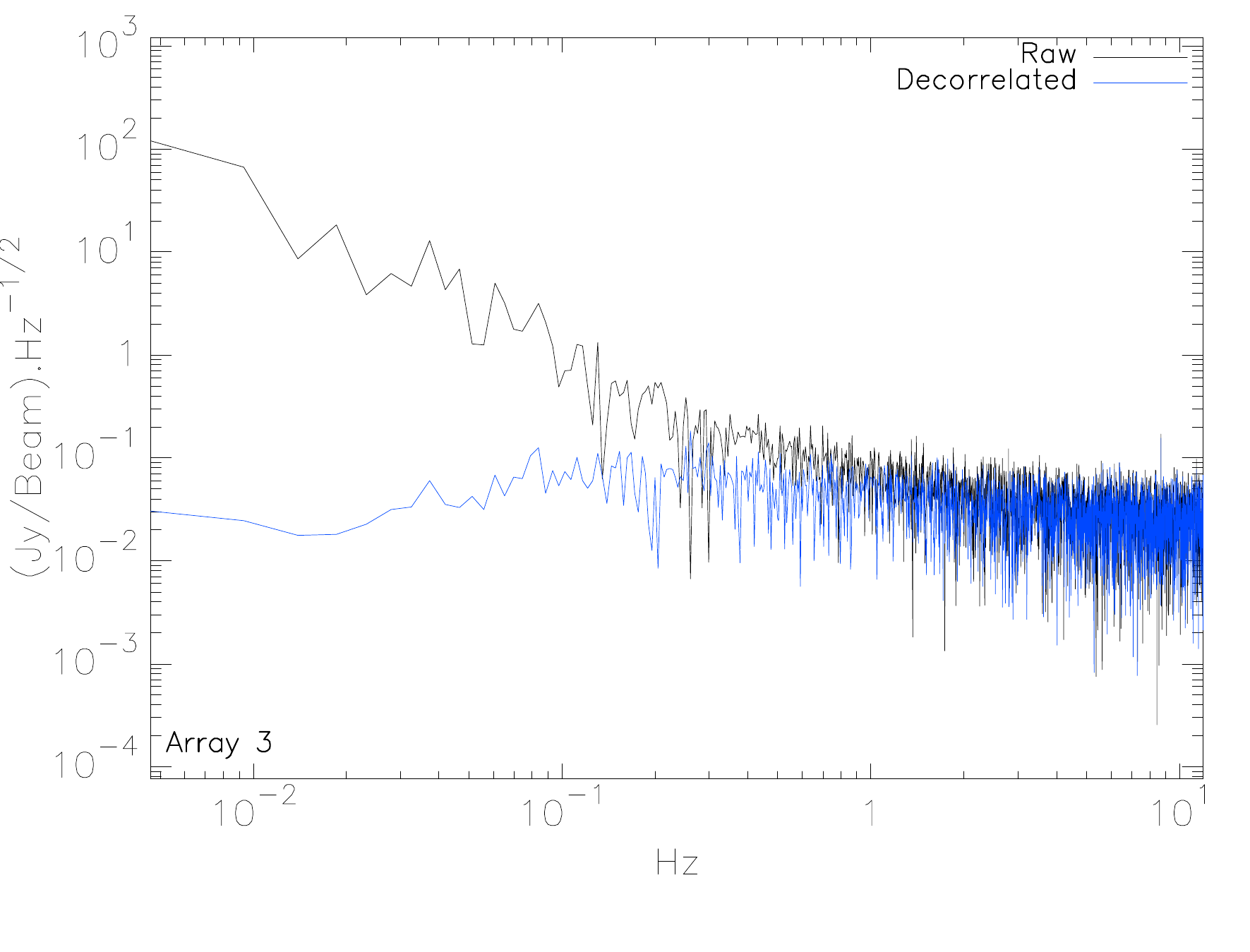}     
    \caption{(Colour online) From top to bottom power spectra of the NIKA2 time ordered data before (black) and after (blue) subtraction of atmospheric fluctuations, which show-up at frequencies below 1 Hz.}
         \label{fig:noisespec}
\end{figure}

\begin{figure}[h]
   \centering(
   \includegraphics[width=.75\linewidth]{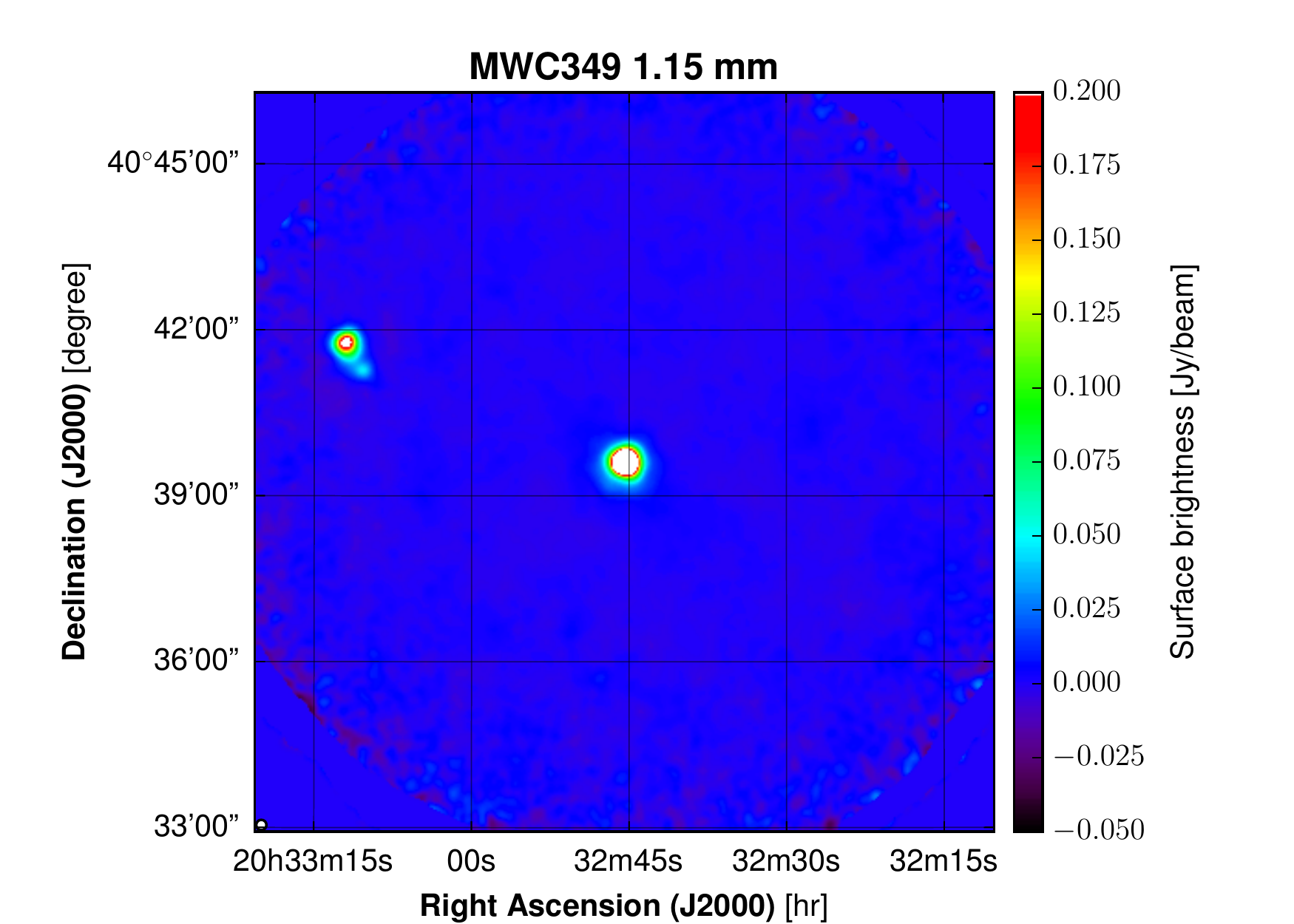}
    \includegraphics[width=.75\linewidth]{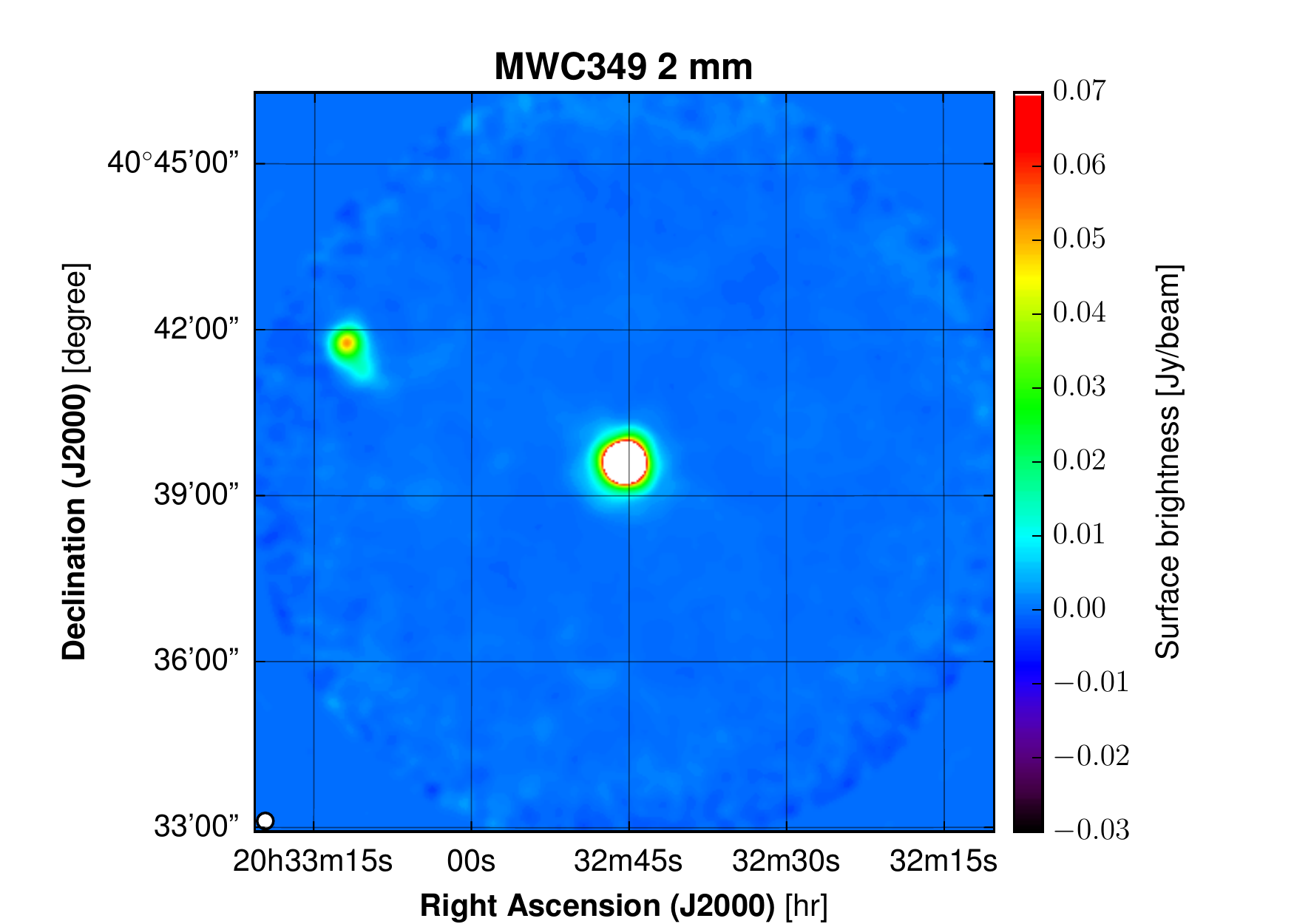} 
      \caption{(Colour online) Maps at 260~GHz (top) and 150~GHz (bottom) centered in MWC349. Details on the observed sources are given in the text. The contours in these maps indicate signal-to-noise ratios of 5, 7 and 10. The FWHM are given in the lower left corners.
         \label{fig_compact_sources1}}
\end{figure}

\begin{figure}[h]
   \centering(
        \includegraphics[width=.75\linewidth]{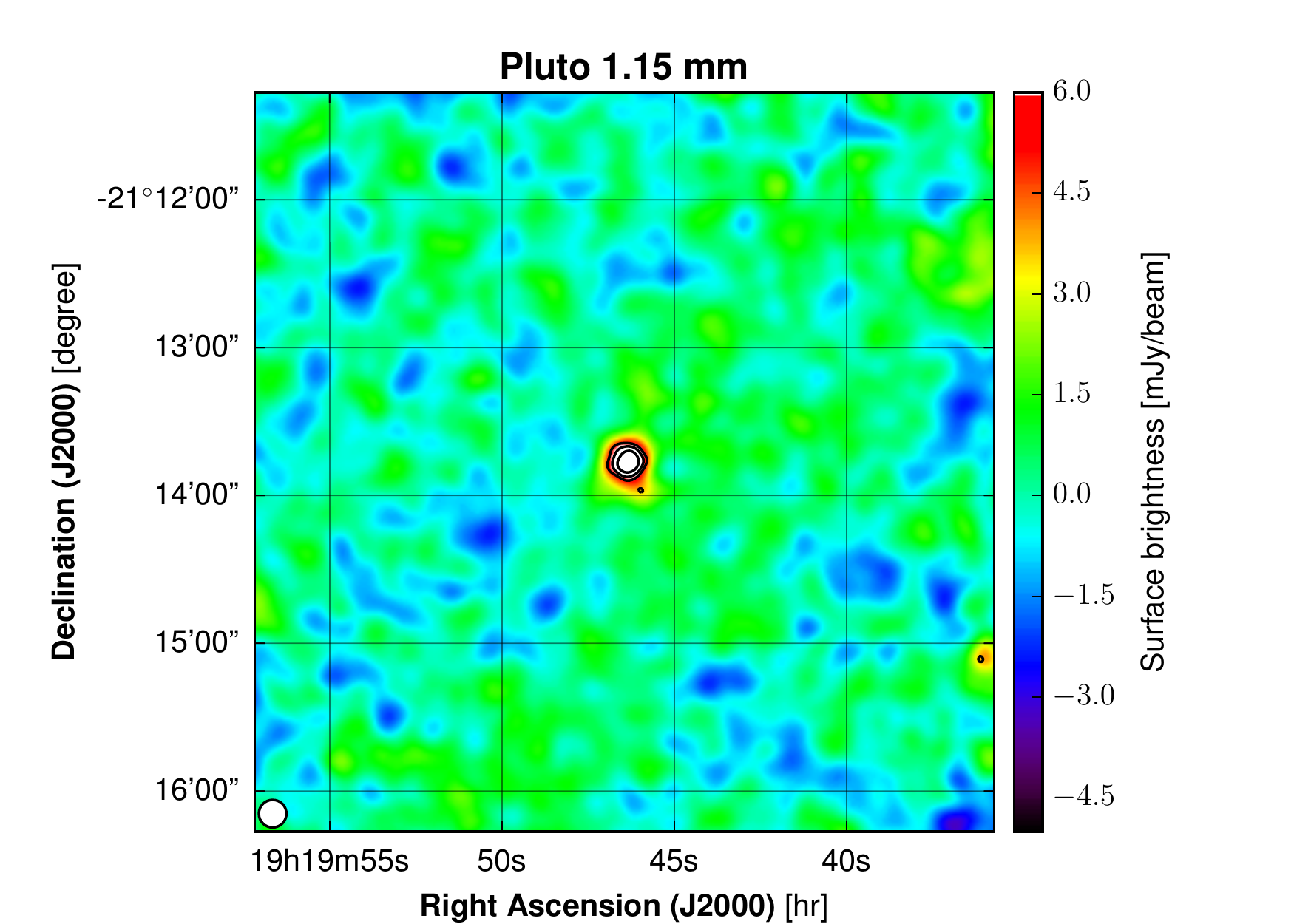}
    \includegraphics[width=.75\linewidth]{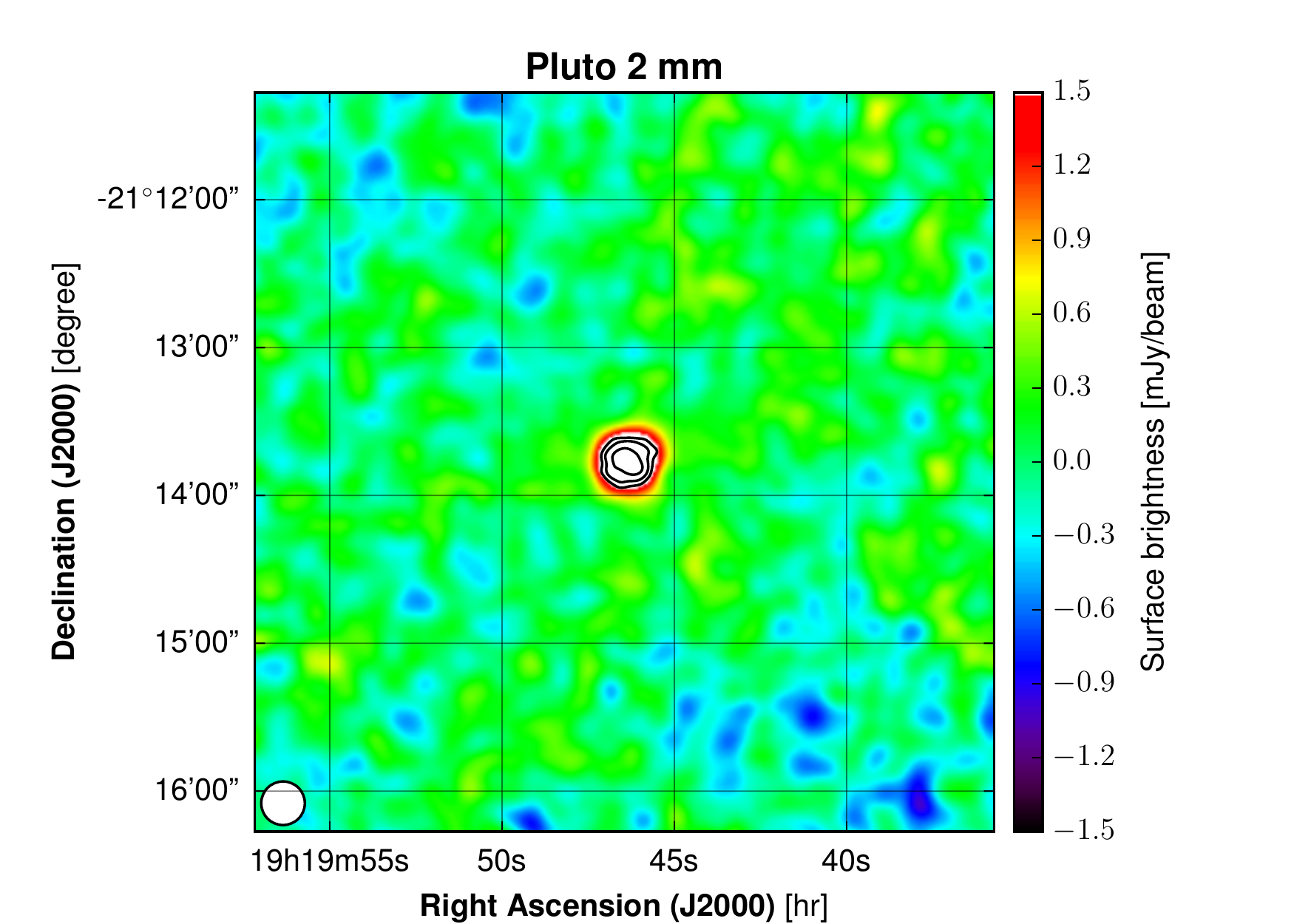}
      \caption{(Colour online) Maps at 260~GHz (top) and 150~GHz (bottom) of the Pluto and Charon planetary system. The contours in these maps indicate signal-to-noise ratios of 5, 7 and 10. The FWHM are given in the lower left corners.
         \label{fig_compact_sources2}}
\end{figure}

\subsection{Noise and sensitivity}
\label{Noise and sensitivity}

We have investigated the noise properties and sensitivity of NIKA2 in various atmospheric conditions and for various types of sources including both faint and bright ones. For each observation scan the raw data were corrected for atmospheric fluctuations, which are seen as a common mode by all or most of the detectors. These corrected data are then projected into maps.

Figure~\ref{fig:noisespec} shows an example, in typical weather conditions $\tau_{260~GHz} < 0.3$, of the power spectrum of the NIKA2 time ordered data before (black) and after (blue) subtraction of the atmospheric fluctuations.
We observe that even in the case of good weather conditions the signal is dominated by atmospheric fluctuations, in particular at small frequencies, giving a 1/f-like spectrum. After atmospheric subtraction and proper filtering, the spectrum is flatter. A detailed study of the residual correlated noise, in both the time-ordered data and maps, will be given in the companion papers \cite{commissioning,pipeline}.


The NEFD is routinely estimated on those maps for each individual scan from the measured flux uncertainties across the map. Using the NEFD we obtain the mapping speed as $m_{s} = \epsilon \cdot FOV / NEFD^2$ where $FOV$ is the field-of-view solid angle and  $\epsilon$ is the fraction of used detectors. We stress that the NEFD values assume the spectral energy distribution (SED) of the primary calibrators.

From our analysis we find that the measured NEFDs per scan are consistent with being background dominated both for the 150 and 260~GHz NIKA2 channels. We observe some residual correlated noise in the per scan maps mainly due to residual atmospheric contamination. However, when averaging across scans the noise evolves consistently with the square root of the time of observation over more than 3 hours of observations. 
We achieved average sensitivities, extrapolated at $\tau = 0$ from a fit (to be easily comparable to other instruments) of 8 and 33~mJy$\cdot\textrm{s}^{1/2}$ at 150 and 260~GHz, respectively. The fit versus tau was performed based on a dataset including observations made under airmasses in the range from 0.05 to 0.85. These NEFDs correspond to mapping speeds of around 1350 and 73 arcmin$^2$/hr/mJy$^2$. These values refer to the average over many scans taken under different observing conditions. They are conservative with respect to the values obtained in the best scans under good weather conditions, that is, 2~mm PWV and elevation $\delta = 60^{\circ}$.
We have thus gained an order of magnitude mapping speed over the previous generation of instruments at the 30-meter telescope (i.e. NIKA and GISMO operating at 150~GHz, NIKA and MAMBO2 operating at 260~GHz).

\subsection{Summary of performances}
We present in Table~\ref{sumperf} a summary of the main characteristics and
performance of the NIKA2 instrument for the February commissioning
campaign. From this table we conclude that NIKA2 behaves better than the
initial goals at 150~GHz, and is compatible with the specifications at 260~GHz
(\cite{Calvo2016}). The NEFD sensitivity at 260~GHz is limited by sky noise
decorrelation techniques (\cite{commissioning, pipeline}) and a still unidentified 
optical problem reducing considerably the illumination on the array A1 (260~GHz-H). 
This issue is under investigation and will be addressed in a forthcoming refurbishment. 

\begin{table*}[t]
  \centering
  \caption{Summary of the principle characteristics and performance of the NIKA2 instrument. \label{sumperf}}
  \begin{tabular}{|c|c|c|c|c|}
    \hline
        Channel & \multicolumn{3}{|c|}{260 GHz} & 150 GHz \\
            & \multicolumn{3}{|c|}{1.15 mm}     &  2 mm \\ 
    \hline
    Arrays & A1 & A3  & A1\&3 & A2 \\
    \hline
    Number of designed detectors       & 1140      &  1140    &    &    616      \\
    Number of valid detectors\tablefootmark{1}     &  952      &   961    &   &    553      \\ 
    Number of used detectors\tablefootmark{2}     &  660 - 830      &   725 -780    &   &    430 - 480      \\ 
    \hline
    FOV diameter [arcmin]     &   6.5              &  6.5              &   6.5        &    6.5        \\
    FWHM [arcsec]             &   $11.3 \pm 0.2$   &  $11.2 \pm 0.2$  &   $11.2 \pm 0.1$           &  $17.7 \pm 0.1$ \\      
    Beam efficiency\tablefootmark{3} [\% ]   & $55 \pm 5$  &  $53 \pm 5$  &  $60 \pm 6$        &     $75 \pm 5$ \\
    \hline 
    rms calibration error [\%]            & 4.5  & 6.6  &   & 5 \\
    \hline
    Model absolute calibration uncertainty [\%] &  \multicolumn{4}{|c|}{5} \\
    \hline
    RMS pointing error    [arcsec]    & \multicolumn{4}{|c|}{$<3$} \\
    \hline
    NEFD [mJy.s$^{1/2}$] \tablefootmark{4}           &    &     & 33$\pm$2      & 8$\pm$1  \\
    Mapping speed [arcmin$^2$/h/mJy$^2$] \tablefootmark{5} &   &   & 67 - 78  & 1288 - 1440 \\
    \hline 
  \end{tabular}
  \tablefoot{
    \tablefoottext{1}{Number of detectors that are valid at least for two different beam map scans.}
    \tablefoottext{2}{Number of detectors used in the scientific analysis after stringent selection (\cite{commissioning, pipeline}).}
    \tablefoottext{3}{Ratio between the main beam power and the total beam power up to a radius of 250$^{\prime \prime}$}
   \tablefoottext{4}{Average NEFD during the February 2017 observation campaign, extrapolated at $\tau = 0$.}   
   \tablefoottext{5}{Average mapping speed during the February 2017 observation campaign, extrapolated at $\tau = 0$.}   
  }
\end{table*}

\section{Illustration of NIKA2 mapping capabilities}
\label{Illustration of NIKA2 mapping capabilities}

\begin{table*}
  \centering
  \caption{NIKA2 measured flux for a selection of sources. Statistical and calibration uncertainties are given. \label{fluxtab}}
\begin{tabular}{|c|c|c|c|c|c|}
\hline
Source         & Observing time [hours]\tablefootmark{a} &  A1 Flux [mJy]  & A3 Flux [mJy] & 260~GHz Flux [mJy]  &  150~GHz Flux [mJy] \\
\hline
\hline
MWC349         & 3.44    &  1994$\pm$1.2$\pm$140 & 2048$\pm$1$\pm$143 & 2027$\pm$1$\pm$142 & 1389.5$\pm$0.2$\pm$97\\
SO1 (20h33$^{\prime}$10.416$^{\prime \prime}$, +40$^{\circ}$41$^{\prime}$15$^{\prime \prime}$ & 3.44 &&& 78$\pm$1$\pm$7&29.7$\pm$0.3$\pm$2 \\
SO2 (20h33$^{\prime}$11.928$^{\prime\prime}$, +40$^{\circ}$41$^{\prime}$44$^{\prime \prime}$& 3.44&&&396$\pm$1$\pm$28&93.8$\pm$0.4$\pm$7 \\
\hline
Pluto-Charon     & 1.44  & 15.8$\pm$1.6$\pm$1.1   & 15.4$\pm$1.2$\pm$1.1 &  15.5$\pm$1.0$\pm$1.1 & 4.8$\pm$0.2$\pm$0.3 \\
\hline
\end{tabular}
  \tablefoot{
    \tablefoottext{a}{Observing time is on-source time excluding time for slewing, pointing, and focussing.} 
}
\end{table*}


During commissioning and the science verification phase we observed several compact and extended sources in order to check the NIKA2 mapping capabilities. Here we concentrate on two sources to illustrate the main advantages of NIKA2 with respect to previous experiments. A more detailed description of the sources observed will be given in companion papers (\cite{commissioning,pipeline}).

In Fig.~\ref{fig_compact_sources1}, we present 260 (top) and 150 (bottom) GHz NIKA2 maps of the star system MWC349, which is a well known secondary calibrator at millimetre wavelengths. MWC349 was systematically observed at different elevations and in different weather conditions during the February and April, 2017, commissioning campaigns to monitor the stability of the calibration. In the Figure, we present the averaged map obtained using all scans taken during the February campaign for a total integration time of 3.44 hours. We clearly observe MWC349 in the centre of the two maps with high significance. The measured MWC349 fluxes are given in Table~\ref{fluxtab}. We observe two other sources at the edges of the maps, which we refer to as SO1 and SO2. Details on the position and fluxes of the sources are given in Table~\ref{fluxtab}. The SO2 source corresponds most probably to the radio-millimetre source BGPS G079.721+00.427, which was detected by Bolocam in their Galactic plane survey (\cite{Rosolowsky2010}). In the Bolocam observations, the SO1 and SO2 sources cannot be distinguished from each other. Furthermore, the SO1 source might correspond to  KMH2014 J203310.31+404118.72 (\cite{Kryukova2014}), a young stellar object, but the identification is not sufficiently secure.
From these observations we conclude that the large FOV and high sensitivity of NIKA2 translate into a large mapping speed that allows us to cover a large sky area with the possibility of observing, and/or discovering, various sources simultaneously. 

To test the ability of NIKA2 to detect faint sources, we performed a concerted series of observations of Pluto and Charon.
During these observations the atmospheric opacity was stable: about 0.25 at 260~GHz. The maps are shown in Fig.~\ref{fig_compact_sources2}. For this paper, we concentrate only on the central region of the maps where we observe a significant detection of the Pluton and Charon planetary system. To illustrate this, on the image, we have also superimposed  signal-to-noise ratio contours at values of 5, 7, and 10. The fluxes and uncertainties of the Pluto and Charon system for the three NIKA2 array observations are given in Table~\ref{fluxtab}. With an observation time of 1.44 hours we reach 1 and 0.3 mJy (1-$\sigma$) at 260~GHz (1.15~mm) and 150~GHz (2~mm), respectively, over a 150~arcmin$^2$ map. 

These results illustrate the high sensitivity of NIKA2, with which mJy sources can be detected in less than one hour.

%
%

\section{Conclusions and future plans}

The NIKA2 instrument has been permanently installed at the 30-meter IRAM telescope since September, 2015. A first technical upgrade was achieved in September 2016. During this upgrade, we replaced the dichroic, changed the 150\,GHz array and replaced most of the smooth lenses with anti-reflecting-coated ones. A number of commissioning observational runs have been achieved since the first light, that is, October, 2015. In the present paper we provided a general overview of the instrument, and show the main results obtained during the commissioning campaigns. 

The performance of the instrument, in terms of sensitivity, surpasses the ambitious goals at 150\,GHz, and is, even per beam, better than previous generation instruments at 260\,GHz. Building on this base, NIKA2 has been opened, in April 2017, to science verification observations, and in October 2017 to general science observations. We are preparing a first purely astrophysical publication centred on high-quality mapping of the high-redshift (z = 0.58) galaxy cluster PSZ2 G144.83+25.11 via the SZ effect (\cite{Ruppin2017}). The instrument is offered, under IRAM responsibility, to a larger community. In parallel, we have entered a phase of commissioning of the polarisation module of NIKA2 following the work performed in NIKA (\cite{Ritacco2017}). 

NIKA2, thanks to its versatile design and to the KID technology adopted, will be upgraded during its lifetime. There are several possible upgrades that we are considering: widening the 260\,GHz channels band in order to match the "1\,mm atmospheric window" in very good observing conditions, adding a third band, reducing the pixel size, adding a polarised channel at 150\,GHz, increasing the illumination of the primary mirror and others. 


\begin{acknowledgements}
We would like to thank the IRAM team in Spain for their work leading NIKA2 towards success; in particular Gregorio Galvez, Pablo Garcia, Israel Hermelo, Dave John, Hans Ungerechts, Salvador Sanchez, Pablo Mellado, Miguel Mu\~noz, Francesco Pierfederici, Juan Pe\~nalver. On top of that, we also would like to thank the remaining IRAM Granada staff for the outstanding support before, during and after the observations. In particular, we thank the telescope operators and the logistics and administration groups. We acknowledge the crucial contributions of the technological groups in N\'eel, LPSC and IRAM Grenoble, and in particular: A. Barbier, E. Barria, D. Billon-Pierron, G. Bosson, J.-L. Bouly, J. Bouvier, G. Bres, P. Chantib, G. Donnier-Valentin,  O. Exshaw, T. Gandit, G. Garde, C. Geraci, A. Gerardin, C. Guttin, C. Hoarau, M. Grollier, C. Li, J. Menu, J.L. Mocellin, E. Perbet, N. Ponchant, G. Pont, H. Rodenas, S. Roni, S. Roudier, J.P. Scordilis, O. Tissot, D. Tourres, C. Vescovi, A.J. Vialle. Their technical and scientific skills, as well as human qualities, represent our main boost.  The NIKA2 contracts have been administrated by the laboratories involved. We acknowledge the contribution of P. Poirier, M. Berard, C. Bartoli, D. Magrez, and F. Vidale, among others. We enjoy frequent fundamental physics discussions concerning superconducting devices with Florence Levy-Bertrand, Olivier Dupr\'e, Thierry Klein, Olivier Buisson and other colleagues at the Institut N\'eel. This work has been mainly funded by the ANR under the contracts "MKIDS", "NIKA", ANR-15-CE31-0017 and LabEx "FOCUS" ANR-11-LABX-0013. NIKA2 has benefited from the support of the European Research Council Advanced Grant ORISTARS under the European Union's Seventh Framework Programme (Grant Agreement no. 291294). We acknowledge funding from the ENIGMASS French LabEx, the CNES post-doctoral fellowship program, the CNES doctoral fellowship program and the FOCUS French LabEx doctoral fellowship program. GL, AB, HA and NP acknowledge financial support from the Programme National de Cosmologie and Galaxies (PNCG) funded by CNRS-CEA-CNES, from the ANR under the contract ANR-15-CE31-0017, the OCEVU Labex (ANR-11-LABX-0060) and the AMIDEX project (ANR-11-IDEX-0001-02) funded by the Investissements d'Avenir program managed by the ANR.

\end{acknowledgements}

%
%


\end{document}